\documentclass{article}

\usepackage{arxiv}
\usepackage[title]{appendix}%
\usepackage[utf8]{inputenc} 
\usepackage[T1]{fontenc}    
\usepackage{hyperref}       
\usepackage{url}            
\usepackage{amsmath,amssymb,amsfonts}%
\usepackage{amsthm}%
\usepackage{mathrsfs}%
\usepackage{bm}
\usepackage{setspace}
\usepackage{float}
\usepackage{multirow}
\usepackage{graphicx}
\usepackage{natbib}

\title{Interior transit orbits in the planar bicircular restricted four-body problem: classification and application}

\author{ \href{https://orcid.org/0009-0001-5111-9779}{\includegraphics[scale=0.06]{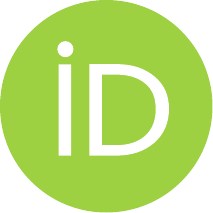}\hspace{1mm}Shuyue Fu}\\
	School of Astronautics\\
	Beihang University\\
	Beijing, China, 100191 \\
	\texttt{fushuyue@buaa.edu.cn} \\
	\And
	{\hspace{1mm}Di Wu} \\
	School of Aerospace Engineering\\
	Tsinghua University\\
	Beijing, China, 100383 \\
	\texttt{wu-d@tsinghua.edu.cn} \\
    \And
	{\hspace{1mm}Shengping Gong$^{*}$}  \\
	School of Astronautics\\
	Beihang University\\
	Beijing, China, 100191 \\
	\texttt{gongsp@buaa.edu.cn} \\
}

\renewcommand{\shorttitle}{\textit{Interior Transit Orbit Analysis in PBCR4BP}}

\hypersetup{
pdftitle={A template for the arxiv style},
pdfsubject={q-bio.NC, q-bio.QM},
pdfauthor={Shuyue Fu, Di Wu, Shengping Gong},
pdfkeywords={Planar bicircular restricted four-body problem, Lagrangian coherent structure, classification of interior transit orbits, low-energy transfer},
}
\begin{document}
\maketitle

\begin{abstract}
Low-energy transfers are advantageous for lunar exploration missions due to low fuel consumption and extended launch periods. This paper is devoted to the classification of interior transit orbits and their application on low-energy transfer in the Sun-Earth/Moon planar bicircular restricted four-body problem (PBCR4BP). First, the Lagrangian coherent structures (LCSs) are introduced to generate the interior transit orbits. The number of periapses about the Moon is selected as the classification parameter and mapped into the LCSs, achieving clear classification boundaries. Then, the evolution laws of the classifications with respect to energy and solar gravity perturbation are discussed and summarized. Construction strategies for low-energy transfer are proposed based on the classifications and their evolution laws. Numerical simulation of the transfer trajectories verifies the effectiveness of the proposed strategies. The dynamical behaviors and transfer characteristics of transit orbits and their families are revealed, and a direct link between transit orbit families and low-energy transfers is finally established.
\end{abstract}


\section{Introduction}\label{sec1}
Lunar exploration has significantly advanced scientific knowledge by enabling deeper insights into the composition, geology, and potential for life on other planets. With the recent proposal and implementation of a series of missions (e.g., \textit{Chang’e-5}, \textit{Chang’e-6}, \textit{Artemis}) \citep{bib1,bib2,bib4}, there has been increased requirement on designing transfer trajectories with high efficiency and low energy in the cislunar space \citep{bib11}, including Earth-Moon transfer trajectories \citep{bib5,bib8}, lunar free-return trajectories \citep{bib12,bib13}, and cislunar escape trajectories \citep{bib15,bib16}. In particular, low-energy transfers with low fuel consumption and extended launch periods \citep{bib11,bib17} have been widely studied and applied to practical missions (e.g., \textit{Hiten}, \textit{Genesis}, \textit{Danuri}) \citep{bib5,bib19,bib20}. Low-energy transfers are usually associated with transit orbits \citep{bib21,bib23}, which are defined as trajectories passing through the neck regions in the multi-body problems (e.g., Earth-Moon circular restricted three-body problem (CR3BP) \citep{bib24}  and Sun-Earth/Moon bicircular restricted four-body problem (BCR4BP) \citep{bib25}).  Therefore, the low-energy transfer characteristics are closely associated with the dynamical behaviors of transit orbits \citep{bib23}. Studies on transit orbits improve our understanding of low-energy transfer and escape mechanisms, aiding in designing these transfers using natural dynamics instead of direct optimization \citep{bib9,bib23,bib60}.

Current studies on transit orbits have focused on their construction methods and applications in designing transfer trajectories. \citet{bib27} associated transit orbits with invariant manifolds. Their work showed that invariant manifolds of the L1/L2 periodic orbits can be used to separate transit orbits from non-transit orbits in the CR3BP. However, when there is a periodic perturbation in the system (e.g., solar gravity perturbation in the Earth-Moon CR3BP, i.e., Sun-Earth/Moon BCR4BP), the time-dependence dynamics precludes the concept of invariant manifolds, complicating the definition of transit orbits \citep{bib21,bib28}. To address this, almost invariant sets that separate transit and non-transit orbits (i.e., Lagrangian coherent structures (LCSs) \citep{bib6,bib30,bib31}) can be introduced in the non-autonomous BCR4BP. The LCSs are defined as the ridges in the finite time Lyapunov exponent (FTLE) fields \citep{bib33}, and the FTLE fields describe the mapping between the initial states and FTLE values of trajectories propagated at a given time. Initial states inside the LCSs generate transit orbits through forward-time propagation, while those outside result in non-transit orbits.

Even inside the LCSs, transit orbits exhibit different patterns and dynamical behaviors depending on the initial states \citep{bib23,bib60}. \citet{bib34} pointed out that the classification of transit orbits can be achieved based on the passage time of transit orbits. However, this method could benefit from further refinement to clearly define the boundaries between different transit orbit families and to provide a more comprehensive geometric description. Meanwhile, in the BCR4BP, the configurations of the LCSs vary with energy and solar gravity perturbation \citep{bib6}, affecting the transit dynamical behaviors and corresponding transfer characteristics. Current studies inadequately explore the evolution of transit orbit families inside the LCSs with respect to energy and solar gravity perturbations. Understanding and using these laws will further aid in the construction of low-energy transfers. Therefore, in this paper, based on the LCSs in the Sun-Earth/Moon planar BCR4BP (PBCR4BP), we investigate the classifications of transit orbits associated with L1 region, i.e., interior transit orbits, and analyze the evolution laws of transit orbit families and their transfer characteristics with respect to energy and solar gravity perturbation. Then, the classifications and their evolution laws are applied to constructing low-energy transfers.

Different from Waalkens et al.’s classification method \citep{bib34} for transit orbits based on passage time, this paper introduces the number of periapses about the Moon as a classification parameter for transit orbits, revealing the geometric properties of transit orbits \citep{bib35}. The mapping between the number of periapses and the initial states of transit orbits is established based on the LCSs. A clear classification boundary is achieved, and the consistency of the developed classification results with those based on the passage time is verified. The evolution laws of the classifications with respect to energy and solar gravity perturbation are analyzed and applied to the two typical scenarios of low-energy transfers, i.e., bi-impulsive Earth-Moon transfer and cislunar escape. Our contribution bridges the gap in understanding the evolution of transit orbit families with respect to energy and solar gravity perturbation in the PBCR4BP and establishes a direct link between transit orbit families and low-energy transfer characteristics.

The rest of this paper is organized as follows. The dynamics of the Sun-Earth/Moon PBCR4BP and LCSs are detailed in Section \ref{sec2}. Section \ref{sec3} presents the generation method of transit orbits and the corresponding characteristic parameters. In Section \ref{sec4}, the classification of transit orbits based on the LCSs is presented. The evolution laws of transit orbit families with respect to energy and solar gravity perturbation are analyzed and discussed. Then, the application of the classification on the two scenarios of low-energy transfers is mentioned in Section \ref{sec5}. Finally, Section \ref{sec6} shows the conclusions of this paper.

\section{Dynamical model}\label{sec2}
The dynamical model for this study is presented in this section, including the PBCR4BP and Lagrangian coherent structure (LCS). Based on the LCSs, interior transit orbits are constructed and investigated.
\subsection{Dynamics of the PBCR4BP}\label{subsec2.1}
\begin{figure}[!htb]
\centering
\includegraphics[width=0.9\textwidth]{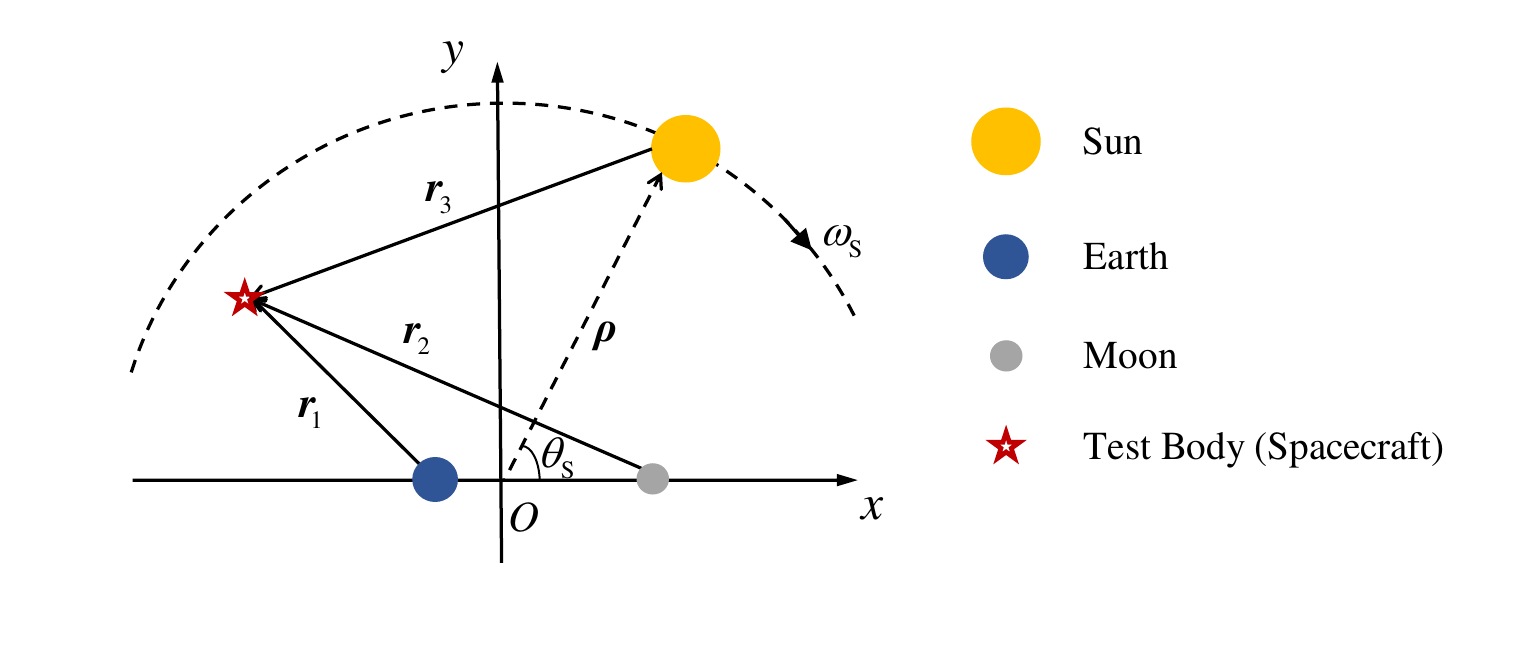}
\caption{Sun-Earth/Moon PBCR4BP in the Earth-Moon rotating frame.}\label{fig1}
\end{figure}

The Sun-Earth/Moon PBCR4BP \citep{bib25} is adopted in this paper where the Sun, Earth, Moon, and test body (spacecraft) move on the same plane. The Earth-Moon barycenter is in the circular orbit about the Sun, while the Earth and Moon are in the circular orbits about their barycenter. The test body (spacecraft) is considered as a massless particle and dominated by the gravity forces from the Sun, Earth, and Moon. The Sun-Earth/Moon PBCR4BP can be considered as a simplified model of the real Sun-Earth/Moon system. As shown in Fig. \ref{fig1}, the Earth-Moon rotating frame \citep{bib6,bib8,bib11} is adopted, and its origin is the Earth-Moon barycenter. The \textit{x} axis is directed from the Earth to the Moon; the \textit{y} axis is perpendicular to the \textit{x} axis and determined by the right-hand rule.

The dimensionless units are set as follows \citep{bib8,bib11}: the length unit (LU) is the Earth-Moon distance; the mass unit (MU) is the combined mass of the Earth and Moon; the time unit (TU) is ${\text{TU}}={T_\text{EM}}/{2\pi}$, where ${T_{{\text{EM}}}}$ is the period of the motion of the Earth and Moon around their barycenter. With these dimensionless units, the equations of the PBCR4BP can be written as:

\begin{equation}
\left[ {\begin{array}{*{20}{c}}
{\begin{array}{*{20}{c}}
{\dot x}\\
{\dot y}
\end{array}}\\
{\begin{array}{*{20}{c}}
{\dot u}\\
{\dot v}
\end{array}}
\end{array}} \right] = \left[ {\begin{array}{*{20}{c}}
{\begin{array}{*{20}{c}}
u\\
v
\end{array}}\\
{\begin{array}{*{20}{c}}
{2v + \frac{{\partial {\Omega _4}}}{{\partial x}}}\\
{ - 2u + \frac{{\partial {\Omega _4}}}{{\partial y}}}
\end{array}}
\end{array}} \right]\label{eq1}
\end{equation}

\begin{equation}
{\Omega _4} = \frac{1}{2}\left[ {{x^2} + {y^2} + \mu \left( {1 - \mu } \right)} \right] + \frac{{1 - \mu }}{{{r_1}}} + \frac{\mu }{{{r_2}}} + \frac{{{\mu_{\text{S}}}}}{{{r_3}}} - \frac{{{\mu_{\text{S}}}}}{{{\rho ^2}}}\left( {x\cos {\theta _{\text{S}}} + y\sin {\theta _{\text{S}}}} \right)\label{eq2}
\end{equation}
where $\bm{X} = \left[ x, y, u, v\right]^{\text{T}}$ is the orbital state, and ${\Omega _4}$ is the effective potential of the PBCR4BP, $\mu $ is the mass parameter expressed as $\mu  = {m_{\text{M}}}/\left( {{m_{\text{E}}} + {m_{\text{M}}}} \right)$, and $\mu_\text{S}  = {m_{\text{S}}}/\left( {{m_{\text{E}}} + {m_{\text{M}}}} \right)$ denotes the dimensionless mass of the Sun. The parameters ${m_{\text{S}}}$, ${m_{\text{E}}}$ and ${m_{\text{M}}}$ denote the masses of the Sun, Earth, and Moon, respectively. The distance between the test body (spacecraft) and the Earth, the Moon, and the Sun $r_1$, $r_2$, and $r_3$ is expressed as:

\begin{equation}
{r_1} = \sqrt {{{\left( {x + \mu } \right)}^2} + {y^2}} \label{eq3}
\end{equation}

\begin{equation}
{r_2} = \sqrt {{{\left( {x + \mu - 1} \right)}^2} + {y^2}} \label{eq4}
\end{equation}

\begin{equation}
{r_3} = \sqrt {{{\left( {x - \rho \cos {\theta _{\text{S}}}} \right)}^2} + {{\left( {y - \rho \sin {\theta _{\text{S}}}} \right)}^2}}  \label{eq5}
\end{equation}
where the solar gravity perturbation depends on the distance between the Sun and the Earth-Moon barycenter ($\rho $), and the solar phase angle ${\theta _{\text{S}}} = {\theta _{{\text{S0}}}} + {\omega _{\text{S}}}T$. Moreover, ${\theta _{{\text{S0}}}} = {\omega _{\text{S}}}{t_0}$ is the solar phase angle at the initial epoch $t_0$ (i.e., at ${t_0} = 0$, the position of the Sun is located at $\left( {\rho ,{\text{ }}0} \right)$ in the Earth-Moon Rotating Frame), ${\omega _{\text{S}}}$ is the synodic solar phase angular velocity, and $T = t - {t_0}$ denotes the propagation time from the initial epoch $t_0$ to the given epoch \textit{t}. The specific values of the aforementioned physical constants used in the simulation are summarized in Table \ref{tab1} \citep{bib8}. When the trajectories in the PBCR4BP are propagated numerically, the variable step-size, variable order (VSVO) Adams-Bashforth-Moulton algorithm \citep{bib8,bib16} is adopted with absolute and relative tolerances set to $1 \times {10^{ - 13}}$.
\begin{table}[h]
\caption{Physical constants of the PBCR4BP}\label{tab1}%
\centering
\renewcommand{\arraystretch}{1.5}
\begin{tabular}{@{}llll@{}}
\hline
Symbol & Value  & Units & Meaning\\
\hline
$\mu$    & $1.21506683 \times {10^{ - 2}}$   & --  & Earth-Moon mass parameter  \\
$\mu_{\text{S}}$    & $3.28900541 \times {10^5}$   & --  & Dimensionless mass of the Sun  \\
$\rho$    & $3.88811143 \times {10^2}$   & --  & Dimensionless Sun-Earth/Moon distance  \\
$\omega_{\text{S}}$    & $ - 9.25195985 \times {10^{ - 1}}$   & --  & Synodic angular velocity  \\
${T_{{\text{EM}}}}$    & $2.24735067 \times {10^6}$   & s  & Earth-Moon period  \\
$R_{\text{E}}$    & $6378.145$   & km  & Mean Earth’s radius  \\
$R_{\text{M}}$    & $1737.100$   & km  & Mean Moon’s radius  \\
LU    & $3.84405000 \times {10^5}$   & km  & Length unit  \\
TU    & $3.75676968 \times {10^5}$   & s  & Time unit  \\
\hline
\end{tabular}
\end{table}

\begin{figure}[!htb]%
\centering
\includegraphics[width=1\textwidth]{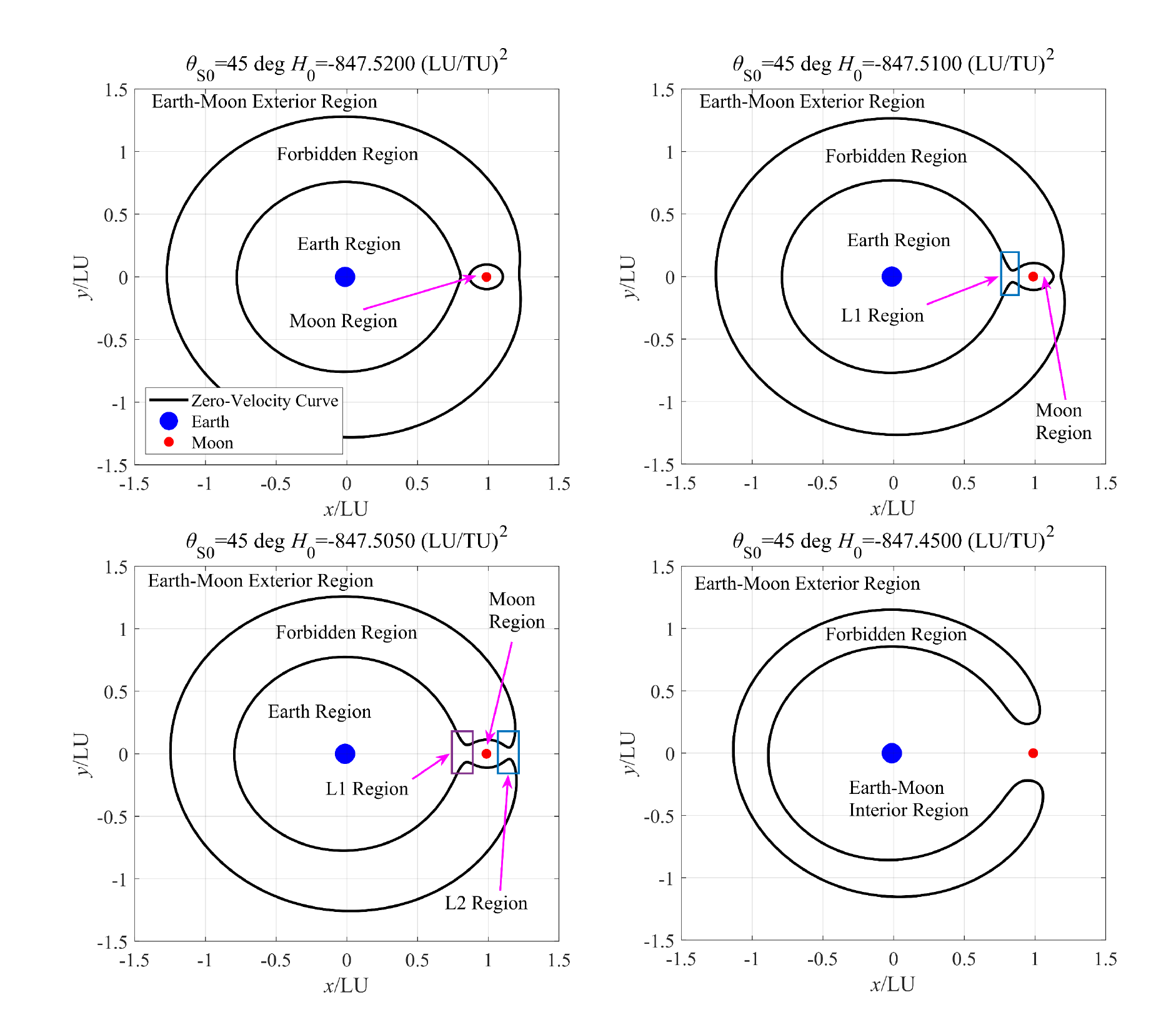}
\caption{Instantaneous Hill regions.}\label{fig2}
\end{figure}

The Hamiltonian indicating the generalized energy of the test body (spacecraft) in the PBCR4BP \citep{bib6} is defined as follows:
\begin{equation}
H = \frac{1}{2}\left( {{u^2} + {v^2}} \right) - \frac{1}{2}\left( {{x^2} + {y^2}} \right) - \frac{{1 - \mu }}{{{r_1}}} - \frac{\mu }{{{r_2}}} - \frac{{{\mu_{\text{S}}}}}{{{r_3}}} + \frac{{{\mu_{\text{S}}}}}{{{\rho ^2}}}x\cos {\theta _{\text{S}}} + \frac{{{\mu_{\text{S}}}}}{{{\rho ^2}}}y\sin {\theta _{\text{S}}} - \frac{1}{2}\mu \left( {1 - \mu } \right) \label{eq6}
\end{equation}
A higher Hamiltonian means a higher level of energy. The Hamiltonian is time-dependent due to the solar gravity perturbation. Following the concept of Hill region \citep{bib27,bib36}, the instantaneous Hill region in the PBCR4BP is defined by the initial Hamiltonian $H_0$ and the initial solar phase angle ${\theta _{{\text{S0}}}}$:
\begin{equation}
S\left( {{H_0},{\theta _{{\text{S0}}}}} \right) = \left\{ {\left( {x,{\text{ }}y} \right)|H\left( {x,{\text{ }}y{\text{, }}u{\text{, }}v{\text{, }}{\theta _{{\text{S0}}}}} \right) \le {H_0},{\text{ }}u = v = 0} \right\} \label{eq7}
\end{equation}
Thus, under a specific initial Hamiltonian $H_0$ and initial solar phase angle ${\theta _{{\text{S0}}}}$, the reachable region can be mainly categorized into the Earth region, the Moon region, the L1 region, the L2 region, and the Earth-Moon exterior region (see Fig. \ref{fig2}) \citep{bib21}. The L1 region and L2 region denote the neck regions. Due to the different patterns of instantaneous Hill regions under different $\left( {{\theta _{{\text{S0}}}},{\text{ }}{H_0}} \right)$, dynamical behaviors of the trajectories vary with $\left( {{\theta _{{\text{S0}}}},{\text{ }}{H_0}} \right)$.

Note that the Hill region parameterized by $\left( {{\theta _{{\text{S0}}}},{\text{ }}{H_0}} \right)$ is time-dependent, so the trajectories in the PBCR4BP may intersect with the instantaneous zero-velocity curves (black curves in Fig. \ref{fig2}) \citep{bib6}. Then, the transit orbits passing through the neck regions are investigated. As the almost invariant sets in the non-autonomous PBCR4BP, LCSs separating the transit orbits from the non-transit orbits \citep{bib6,bib30} are introduced.

\subsection{Lagrangian coherent structure}\label{subsec2.2}
\begin{figure}[!htb]%
\centering
\includegraphics[width=0.9\textwidth]{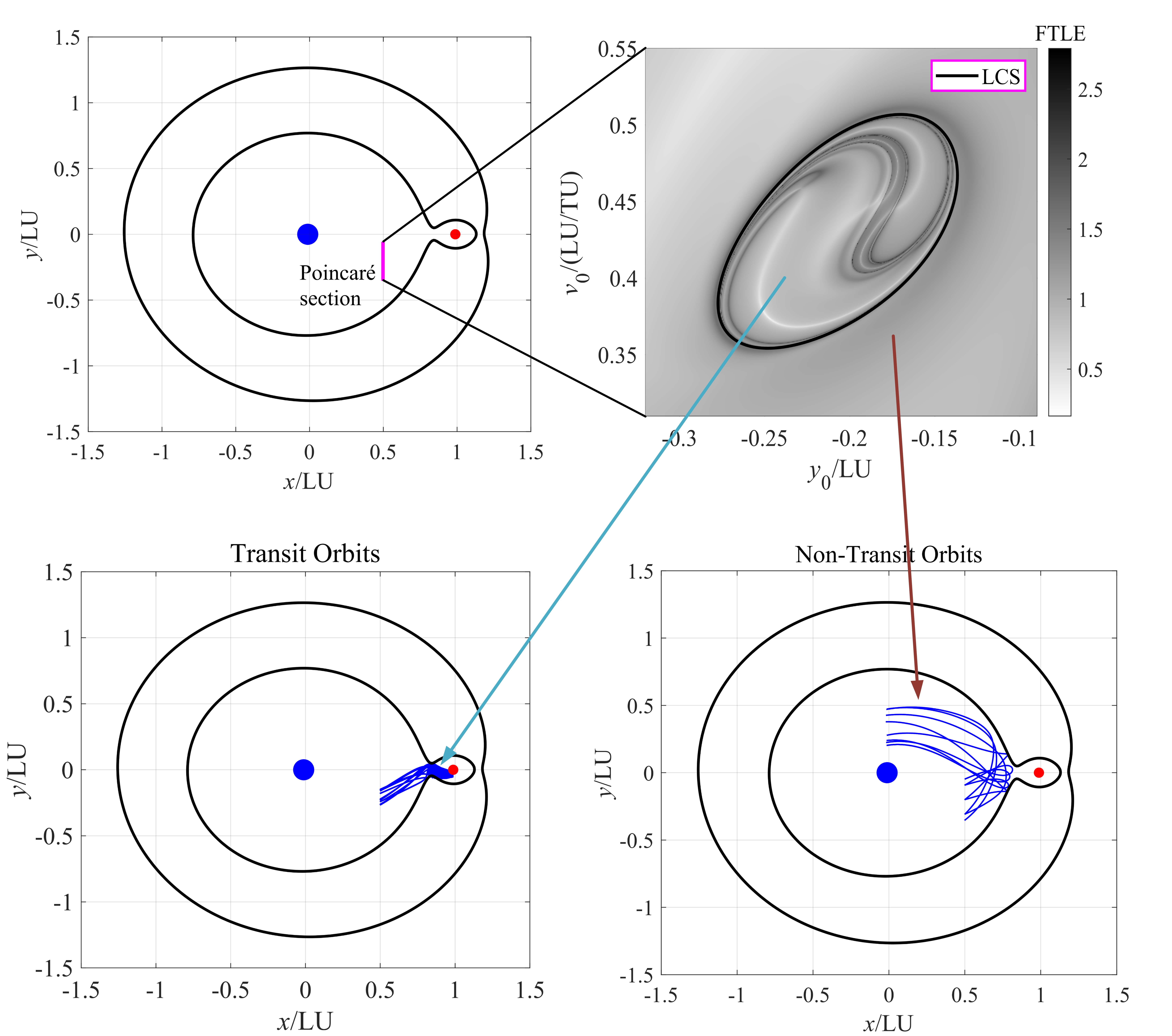}
\caption{The role of the LCS as a separatrix.}\label{fig3}
\end{figure}

In the non-autonomous PBCR4BP, the LCSs are introduced to describe the phase space transport and generate transit orbits \citep{bib6,bib30}. Beforehand, the finite time Lyapunov exponent (FTLE) and FTLE fields are introduced to define the LCSs. To reduce the dimension of the phase space and visualize the FTLE field, the initial Poincaré section $U_0$ (i.e., at the initial epoch $t_0$) in this paper is selected as:
\begin{equation}
{U_0} = \left\{ {\left( {{y_0},{v_0}} \right)|{x_0} = {\text{const}},{\text{ }}{H_0} = {\text{const,}}\text{ }{\theta _{{\text{S0}}}} = {\text{const}}} \right\} \label{eq8}
\end{equation}
where the subscript ‘0’ denotes the quantities associated with $U_0$ at the initial epoch  ${t_0} = {\theta _{{\text{S0}}}}/{\omega _{\text{S}}}$. In the following texts, the $U_0$ is set as ${U_0} = \left\{ {\left( {{y_0},{v_0}} \right)|{x_0} = 0.5\text{ }{\text{ LU}},{\text{ }}{H_0} = {\text{const,}}\text{ }{\theta _{{\text{S0}}}} = {\text{const}}} \right\}$. When $U_0$ is selected, ${u_0} = {\dot x_0}$ is calculated as:
\begin{equation}
u_0=\pm \sqrt{\begin{array}{l}
	-2H_0+{x_0}^2+{y_0}^2-{v_0}^2+\frac{2\left( 1-\mu \right)}{r_{10}}+\frac{2\mu}{r_{20}}+\frac{2\mu_{\text{S}}}{r_{30}}\\
	-\frac{2\mu_{\mathrm{S}}}{\rho ^2}\left( x_0\cos \theta _{\text{S0}}+y_0\sin \theta _{\text{S0}} \right) +\mu \left( 1-\mu \right)\\
\end{array}}
\label{eq9}
\end{equation}
where $\left( {{y_0},{\text{ }}{v_0}} \right)$ are selected from the grid on $U_0$ and the signs of $u_0$ depend on the direction of the orbital propagation. The positive sign of $u_0$ is adopted to obtain the trajectories towards the Moon \citep{bib6,bib30}. Therefore, the set of initial states of trajectories ($M$) is selected on $U_0$, i.e., $\left( {{y_0},{\text{ }}{v_0}} \right)$ under the specific $\left( {{\theta _{{\text{S0}}}},{\text{ }}{H_0}} \right)$. The trajectories are generated from the forward time propagation of $M$ on $U_0$ and the FTLE values associated with these trajectories can be calculated as \citep{bib29}:
\begin{equation}
\sigma \left( {{\bm{X}_0},{\text{ }}{t_{\text{0}}}{\text{, }}T} \right) = \frac{1}{T}\ln {\left( {\sqrt {{\lambda _{\max }}\left( {{{\left( {\nabla \phi _{{t_0}}^{{t_0} + T}\left( {{\bm{X}_0}} \right)} \right)}^{\text{T}}}\nabla \phi _{{t_0}}^{{t_0} + T}\left( {{\bm{X}_0}} \right)} \right)} } \right)_{{\bm{X}_0} \in M}} \label{eq10}
\end{equation}
where ${\lambda _{\max }}\left( {{{\left( {\nabla \phi _{{t_0}}^{{t_0} + T}\left( {{\bm{X}_0}} \right)} \right)}^{\text{T}}}\nabla \phi _{{t_0}}^{{t_0} + T}\left( {{\bm{X}_0}} \right)} \right)$ is the maximum eigenvalue of the Cauchy-Green tensor ${\left( {\nabla \phi _{{t_0}}^{{t_0} + T}\left( {{\bm{X}_0}} \right)} \right)^{\text{T}}}\nabla \phi _{{t_0}}^{{t_0} + T}\left( {{\bm{X}_0}} \right)$, and $\phi _{{t_0}}^{{t_0} + T}:{M} \to {\mathbb{R}^4}{\text{ }}$ is the flow map of the PBCR4BP. The set of FTLE values $M_2$ is obtained, the mapping $\Re :{\text{ }}M \to {M_2}$ is established, and the FTLE field is visualized (see grey region in Fig. \ref{fig3}). The LCSs are defined as the ridges in the FTLE fields \citep{bib32,bib33}, shown as a closed curve on the $\left( {{y_0},{\text{ }}{v_0}} \right)$ plane (see black curve in Fig. \ref{fig3}). The LCS separates the transit orbits (blue trajectories reaching the Moon region in Fig. \ref{fig3}) from the non-transit orbits (blue trajectories bounded back to the Earth region in Fig. \ref{fig3}), i.e., the initial states inside the LCSs (the set of these initial states is denoted as $M_3$) can be propagated forwards in time to obtain transit orbits. 

\begin{figure}[h]%
\centering
\includegraphics[width=1\textwidth]{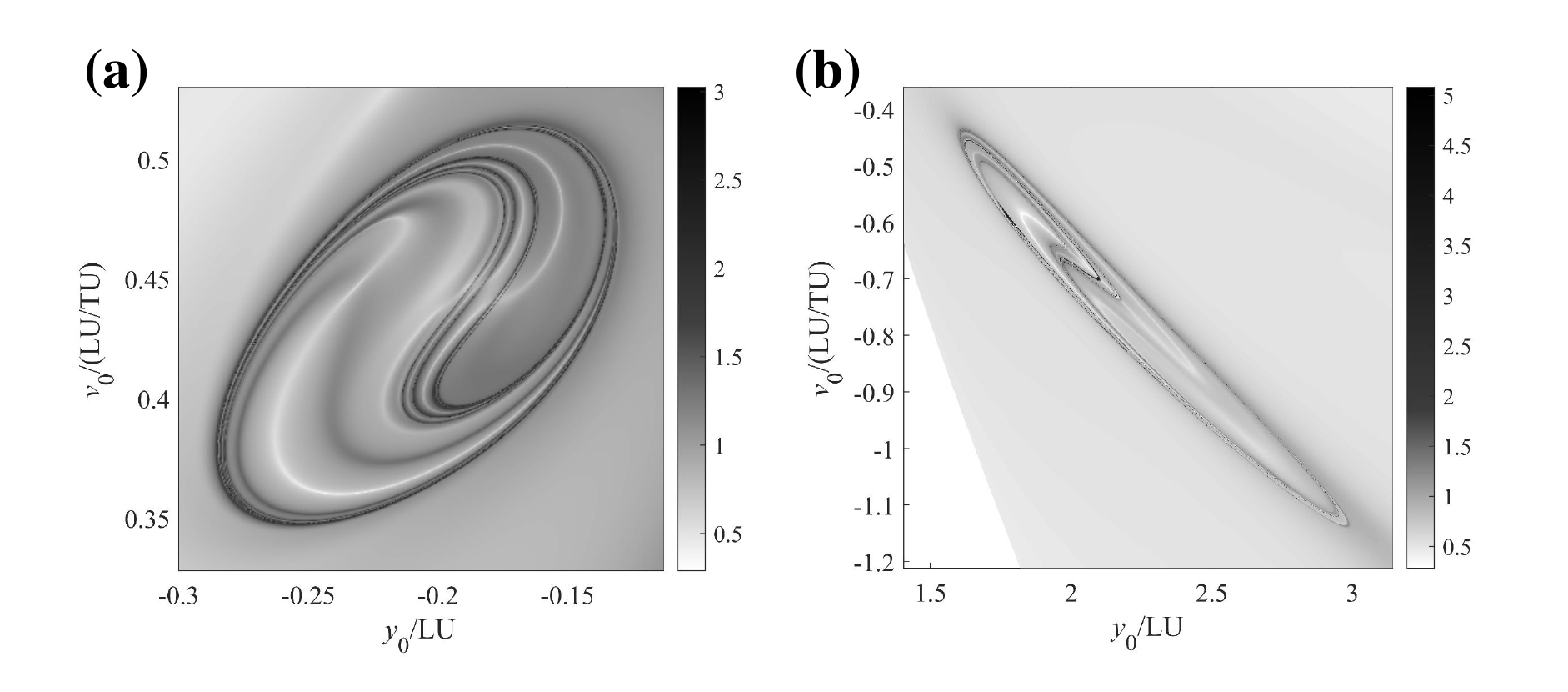}
\caption{Examples of interior-LCSs and exterior-LCSs in the FTLE fields. (a) Interior-LCS; (b) Exterior-LCS.}\label{fig4}
\end{figure}

Similar to the invariant manifolds of the L1/L2 periodic orbits \citep{bib27,bib41} in the autonomous planar circular restricted three-body problem (PCR3BP), there are interior-LCSs associated with the L1 region and exterior-LCSs associated with the L2 region (see examples of interior-LCSs and exterior-LCSs on $U_0$ under the specific $\left( {{\theta _{{\text{S0}}}},{\text{ }}{H_0}} \right)$ in Fig. \ref{fig4}). The LCS presented in Fig. \ref{fig3} is an interior-LCS. Consequentially, interior transit orbits (i.e., trajectories passing through the L1 region) can generated from the forward time propagation of the initial states inside the interior-LCSs while exterior transit orbits (i.e., trajectories passing through the L2 region) can obtained from exterior-LCSs. In this paper, the dynamical behaviors and characteristics of interior transit orbits are focused on and analyzed based on the interior-LCSs parameterized by $\left( {{\theta _{{\text{S0}}}},{\text{ }}{H_0}} \right)$. In the following texts, interior transit orbits and interior-LCSs are abbreviated as transit orbits and LCSs. Even inside the LCSs, the patterns of transit orbits are different \citep{bib23,bib60}, which results in the difference in their characteristics. Therefore, this paper investigates transit orbits in terms of transit orbit families to reveal the link between transit orbit patterns and their transfer characteristics.

\begin{figure}[!htb]%
\centering
\includegraphics[width=0.55\textwidth]{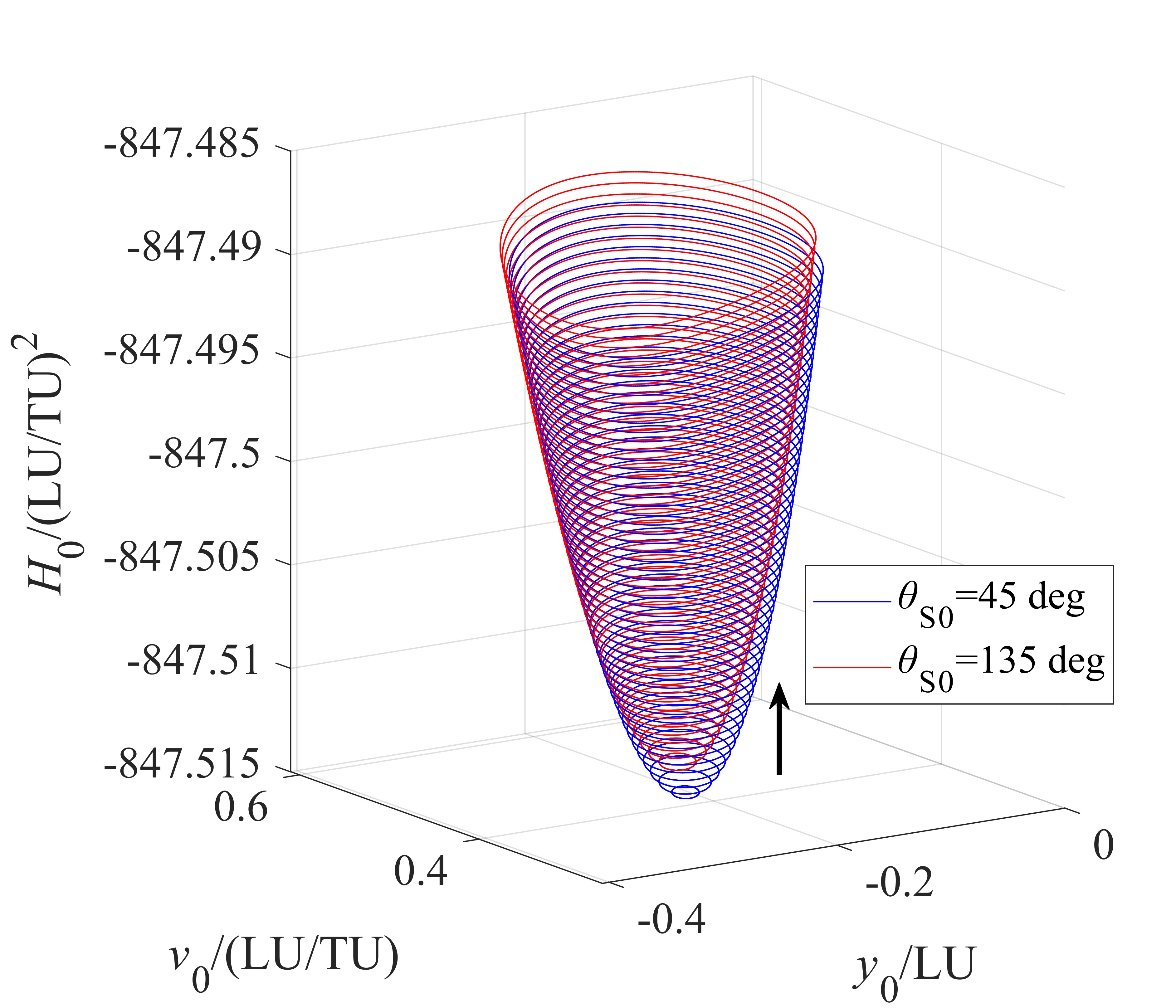}
\caption{The effect of solar gravity perturbation on the LCSs.}\label{fig5}
\end{figure}

As mentioned above, the $U_0$ is set as ${U_0} = \left\{ {\left( {{y_0},{v_0}} \right)|{x_0} = 0.5{\text{ LU}},{\text{ }}{H_0} = {\text{const, }}{\theta _{{\text{S0}}}} = {\text{const}}} \right\}$. Therefore, the LCSs on the $\left( {{y_0},{\text{ }}{v_0}} \right)$ plane are parameterized by $\left( {{\theta _{{\text{S0}}}},{\text{ }}{H_0}} \right)$ to investigate the effect of energy and solar gravity perturbation. At a specific ${\theta _{{\text{S0}}}}$, when $H_0$ on $U_0$ is low enough, the LCS will shrink to one point \citep{bib6}, as shown in Fig. \ref{fig5}. This implies that the existence of LCSs depends on a specific range of $H_0$. When ${\theta _{{\text{S0}}}}$ varies,  Qi et al. \citep{bib6}  pointed out that the configurations of LCSs translate along the  $H_0$  axis (see Fig. \ref{fig5}). In addition, we further find that the configurations of LCSs also translate in the $\left( {{y_0},{\text{ }}{v_0}} \right)$ plane. The parameter pair $\left( {{\theta _{{\text{S0}}}},{\text{ }}{H_0}} \right)$ affects the configurations and distributions of the LCSs \citep{bib6}, and further affects the dynamical behaviors of transit orbits (i.e., the patterns of them) and transfer characteristics of their families. The minimum $H_0$ (${H_{0\min }}$) required for the existence of the LCSs at each ${\theta _{{\text{S0}}}}$ is presented in Table \ref{tab2}. Then, the evolution laws of transit orbits and their families with respect to $\left( {{\theta _{{\text{S0}}}},{\text{ }}{H_0}} \right)$ are discussed and analyzed. Beforehand, the generation method of transit orbits and corresponding parameters are presented in Section \ref{sec3}.

\begin{table}[h]
\renewcommand{\arraystretch}{1.5}
\centering
\caption{${H_{0\min }}$ for LCSs at each ${\theta _{{\text{S0}}}}$}\label{tab2}%
\begin{tabular}{@{}lllll@{}}
\hline
${\theta _{{\text{S}}0}}/\deg $    & 45   & 135  & 225  & 315  \\
\hline

${H_{0\min }}/{\left( {{\text{LU/TU}}} \right)^2}$    &  $-847.5150$  & $-847.5135$  & $-847.5150$  & $-847.5135$  \\
\hline
\end{tabular}

\end{table}

\section{Transit orbit and classification method}\label{sec3}
In this section, the generation method of transit orbits is first presented based on the LCSs. Then, based on the propagation of transit orbits, an important parameter, i.e., the number of periapses about the Moon (\textit{N}) is introduced. This parameter can naturally classify the transit orbits into families with different characteristics. Then, parameters associated with these characteristics are presented. 
\subsection{Generation of transit orbits}\label{subsec3.1}
The step sizes of $\left( {{y_0},{\text{ }}{v_0}} \right)$ grid in the  $M_3$ under the specific $\left( {{\theta _{{\text{S0}}}},{\text{ }}{H_0}} \right)$ are set as 0.0002 LU and 0.0002 ${\left( {{\text{LU/TU}}} \right)^2}$ to generate transit orbits. Based on selected $\left( {{y_0},{\text{ }}{v_0}} \right)$ in $M_3$, the transit orbits are propagated forward in time. The Moon is treated as a particle with ${m_{\text{M}}}$ to ensure successive propagation of the transit orbits within the Moon region. Note that the propagation will stop when the transit orbit impacts the center of the Moon (i.e., singular collision orbits  \citep{bib26}). These singular collision orbits are not the primary focus of this paper. 

The orbital propagation time is set to  $T = 10000{\text{ TU}}$ to make the transit orbits have enough time to pass through the L1/L2 region \citep{bib15}. Propagation stops when the transit orbits pass through the L1/L2 region (i.e. when the transit orbits reach the termination Poincaré sections ${U_1} = \left\{ {\bm{X}|x = 0.7{\text{ LU}},{\text{ }}u < 0} \right\}$ or ${U_2} = \left\{ {\bm{X}|x = 1.2{\text{ LU}},{\text{ }}u > 0} \right\}$). The actual propagation time is denoted as the passage time in the Moon region (${T_{{\text{Passage}}}}$). The trajectories reaching $U_1$ are defined as L1 escape trajectories, while the orbits reaching $U_2$ are defined as L2 escape trajectories \citep{bib9,bib57}. According to Liouville's theorem \citep{bib42}, excluding the singular collision orbits, L1 escape trajectories and L2 escape trajectories consist of transit orbits. When the propagation of transit orbits is finished, a parameter, i.e., the number of periapses about the Moon is introduced to classify the transit orbits.

\subsection{Classification of transit orbits}\label{subsec3.2}
When the propagation of transit orbits is finished, the number of periapses of each trajectory is recorded. The states of periapsis about the Moon satisfy \citep{bib35}:
\begin{equation}
q = \left( {{x_p} + \mu  - 1} \right)\left( {{u_p} - {y_p}} \right) + {y_p}\left( {{v_p} + {x_p} + \mu  - 1} \right) = 0 \label{eq11}
\end{equation}
\begin{equation}
\dot q = \left( {{x_p} + \mu  - 1} \right)\left( {{{\ddot x}_p} - {v_p}} \right) + {u_p}\left( {{u_p} - {y_p}} \right) + {y_p}\left( {{{\ddot y}_p} + {u_p}} \right) + {v_p}\left( {{v_p} + {x_p} + \mu  - 1} \right) > 0 \label{eq12}
\end{equation}
where the subscript ‘\textit{p}’ denotes the quantities associated with the periapsis about the Moon. Equation \eqref{eq11} denotes that the position vector relative to the Moon is normal to the velocity vector in the inertial frame, and the time derivative $\dot q > 0$ geometrically guarantees that the position of the states $\bm{X}_p = \left[ x_p, y_p, u_p, v_p\right]^{\text{T}}$ satisfying Eq. \eqref{eq12} corresponds to a periapsis rather than an apoapsis. In this paper, the number of the periapses at ${T_{{\text{Passage}}}}$ ($N$) is used as the classification parameter. \textit{N} reveals the stability of transit orbits about the Moon, such that a higher $N$ corresponds to a longer ${T_{{\text{Passage}}}}$. Transit orbits with the different $N$ belong to different families. Similarly, transit orbits with the same $N$ but different periapsis distribution (i.e., different patterns shown in \citet{bib23} and \citet{bib60}, which is reflected in the discontinuous distributions in $M_3$) also represent different families.

When the propagation of transit orbits is finished, the value of \textit{N}  for each trajectory is recorded, and the mapping $\Im :{\text{ }}{M_3} \to N$ is established. Then, the classifications of transit orbits under specific $\left( {{\theta _{{\text{S0}}}},{\text{ }}{H_0}} \right)$ are established based on \textit{N} and periapsis distribution.  To quantitatively measure the difference between the periapsis distribution of transit orbits with the same \textit{N} and exactly extract the initial state set of the same family, the density-based spatial clustering of applications with noise (DBSCAN) method \citep{bib35,bib61} is adopted. Clustering is performed for the transit orbits with the same \textit{N} to classify transit orbits with different periapsis distribution into different families. The compressed description vector is set as:

\begin{equation}
{\bm{Y}_{Nj}} = \left[ {{y_0},{\text{ }}{v_0},{\text{ }}{T_{{\text{Passage}}}},{\text{ }}{\bm{X}_{p1}}^{\text{T}},{\text{ }}...,{\text{ }}{\bm{X}_{pi}}^{\text{T}},{\text{ }}...,{\text{ }}{\bm{X}_{pN}}^{\text{T}},{\text{ }}L} \right]\label{eq21}
\end{equation}
where ${\bm{X}_{pi}} = {\left[ {{x_{pi}},{\text{ }}{y_{pi}},{\text{ }}{u_{pi}},{\text{ }}{v_{pi}}} \right]^{\text{T}}}{\text{ }}\left( {i = 1,{\text{ }}...,{\text{ }}N} \right)$ is the orbital state of the \textit{i}-th peripasis of the transit orbit. The subscript ‘\textit{j}’ denotes the compressed description vector of \textit{j}-th trajectory of transit orbits with \textit{N}. The quantity \textit{L} determines whether the transit orbit is an L1 escape trajectory, an L2 escape trajectory, or some other trajectory:
\begin{equation}
L = \left\{ \begin{gathered}
  1{\text{\text{      }L1 escape trajectory}} \hfill \\
  2{\text{\text{      }L2 escape trajectory}} \hfill \\
  0{\text{\text{      }other}} \hfill \\ 
\end{gathered}  \right.\label{eq22}
\end{equation}
The vectors ${\bm{Y}_{Nj}}$ are combined to form the dataset ${\left[ \bm{S} \right]_N}$ and the clustering is performed to extract the initial state (i.e., $(y_0,\text{ }v_0)$) set. After the clustering, the transit orbit families denoted as F11, F12, etc., detailed in Section \ref{sec4}, and their initial state sets are denoted as $M_{\text{F11}}$, $M_{\text{F12}}$, etc. Subsequently, the transfer characteristic parameters of typical families are analyzed to reveal the evolution laws with respect to $\left( {{\theta _{{\text{S0}}}},{\text{ }}{H_0}} \right)$. These parameters include the percentages ($\eta $) \citep{bib15,bib43}, the average ${T_{{\text{Passage}}}}$ ($\left\langle {{T_{{\text{Passage}}}}} \right\rangle $) \citep{bib34} and the ranges of periapsis altitudes ($h_p$) \citep{bib8,bib11}.

The percentage of transit orbit families is expressed as (take family F11 for example) \citep{bib43}:
\begin{equation}
{\eta _{{\text{F}}11}} = \frac{{m\left( {{M_{{\text{F}}11}}} \right)}}{{m\left( {{M_3}} \right)}} = \frac{{\iint_{\left( {{y_0},{\text{ }}{v_0}} \right) \in {M_{{\text{F}}11}}} {\text{ }}{\text{d}}{y_0}{\text{d}}{v_0}}}{{\iint_{\left( {{y_0},{\text{ }}{v_0}} \right) \in {M_3}} {\text{ }}{\text{d}}{y_0}{\text{d}}{v_0}}} \label{eq13}
\end{equation}
where $m\left(  \cdot  \right)$ denotes the measurement of the sets, ${M_{{\rm{F}}11}}$ denotes the initial state set of family F11 (apparently, ${M_{{\rm{F}}11}} \subseteq {M_3}$). The percentages of transit orbit families reveal the variation of family distribution inside the LCSs with $\left( {{\theta _{{\rm{S0}}}},{\rm{ }}{H_0}} \right)$. 

The average  ${T_{{\text{Passage}}}}$ ($\left\langle {{T_{{\text{Passage}}}}} \right\rangle $) of families is expressed as (take family F11 for example) \citep{bib34}:
\begin{equation}
{\left\langle {{T_{{\text{Passage}}}}} \right\rangle _{{\text{F}}11}} = \frac{{\iint_{\left( {{y_0},{\text{ }}{v_0}} \right) \in {M_{{\text{F}}11}}} {{T_{{\text{Passage}}}}{\text{d}}{y_0}{\text{d}}{v_0}}}}{{m\left( {{M_{{\text{F}}11}}} \right)}} = \frac{{\iint_{\left( {{y_0},{\text{ }}{v_0}} \right) \in {M_{{\text{F}}11}}} {{T_{{\text{Passage}}}}{\text{d}}{y_0}{\text{d}}{v_0}}}}{{\iint_{\left( {{y_0},{\text{ }}{v_0}} \right) \in {M_{{\text{F}}11}}} {\text{ }}{\text{d}}{y_0}{\text{d}}{v_0}}} \label{eq14}
\end{equation}
where ${T_{{\text{Passage}}}}$ denotes the passage time associated with trajectories propagated by the initial states $\left( {{y_0},{\text{ }}{v_0}} \right) \in {M_{{\text{F}}11}}$. $\left\langle {{T_{{\text{Passage}}}}} \right\rangle $ is related to the time of flight (TOF) of low-energy transfers constructed from transit orbits and the evaluation of lunar flyby (LF) \citep{bib47,bib48} in low-energy transfers.

The periapsis altitude of transit orbits about the Moon is calculated by \citep{bib8,bib11}:
\begin{equation}
{h_p} = \sqrt {{{\left( {{x_p} + \mu  - 1} \right)}^2} + {y_p}^2}  - {R_{\text{M}}} \label{eq15}
\end{equation}
The periapsis altitude about the Moon of transit orbits is associated with the altitude of lunar insertion orbits in the low-energy transfers. Therefore, $\left\langle {{T_{{\text{Passage}}}}} \right\rangle $ and the ranges of periapsis altitudes are parameters that reveal transfer characteristics. 

\begin{figure}[!htb]%
\centering
\includegraphics[width=0.5\textwidth]{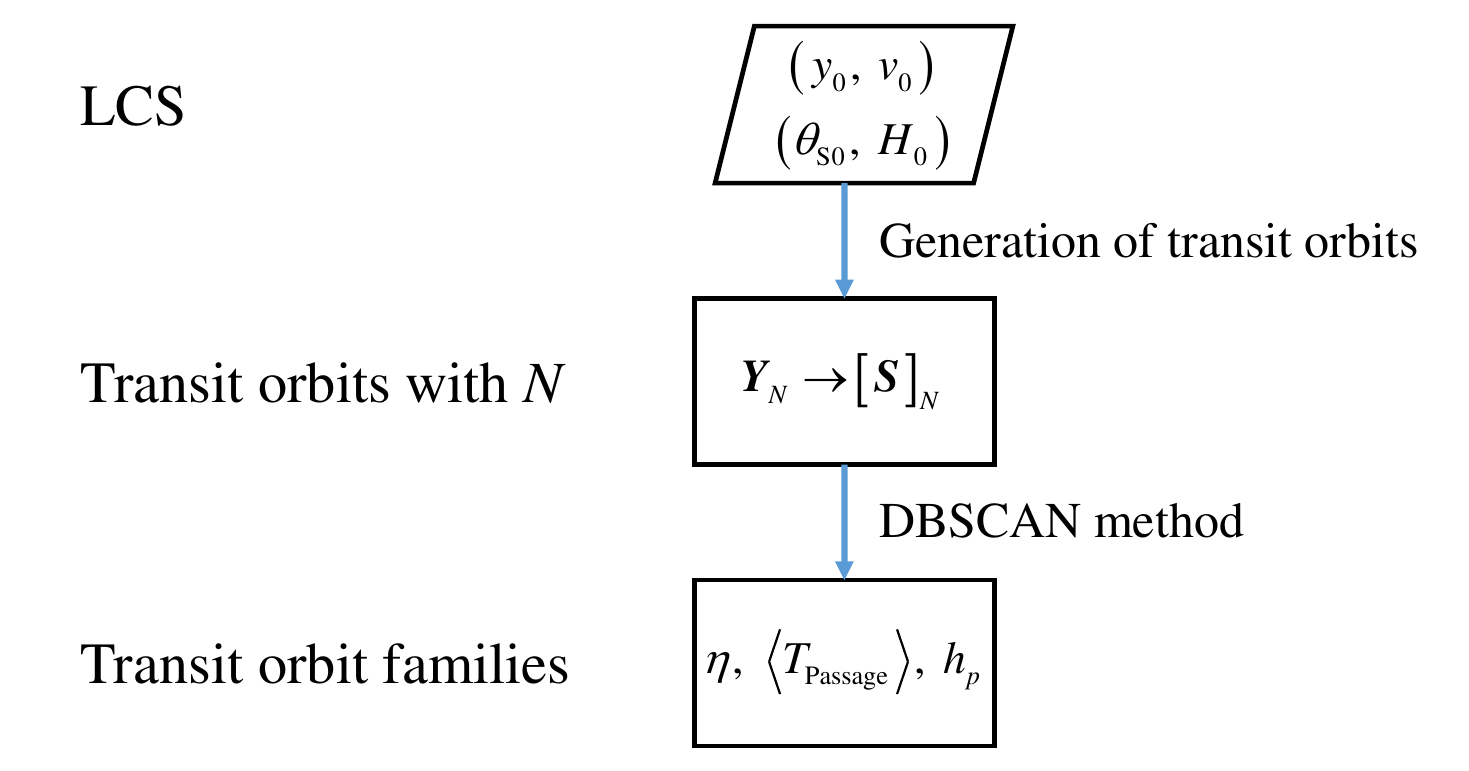}
\caption{Process of classification.}\label{fig31}
\end{figure}

Above all, the process of classification is presented in Fig. \ref{fig31}. Then, the classifications of transit orbits and evolution laws with respect to $\left( {{\theta _{{\text{S0}}}},{\text{ }}{H_0}} \right)$ will be detailed.

\section{Classifications of transit orbits}\label{sec4}
In this section, the classifications of transit orbits are achieved by the mapping $\Im :{\text{ }}{M_3} \to N$ based on the LCSs. First, the global map of classification with a range of $H_0$ at a specific ${\theta _{{\text{S0}}}}$ is presented. Then, the typical maps of classifications are selected to detail the patterns and periapsis distributions of transit orbit families. Similar analyses are subsequently performed for other ${\theta _{{\text{S0}}}}$ cases. Finally, the evolution laws of transfer characteristic parameters for transit orbit families with respect to $\left( {{\theta _{{\text{S0}}}},{\text{ }}{H_0}} \right)$ are discussed and summarized.

\subsection{Case I: $\text{45}{\text{ }}\text{deg} $}\label{subsec4.1}
Our exploration begins with the case at ${\theta _{{\text{S0}}}} = 45{\text{ }}\deg $. Subsequently, the similarities and differences between the classifications at ${\theta _{{\text{S0}}}} = 45{\text{ }}\deg $ and those at other ${\theta _{{\text{S0}}}}$ are discussed (Subsections \ref{subsec4.2}-\ref{subsec4.3}). The existence of LCSs depends on ${H_{0\min }}$, as presented in Table \ref{tab2} \citep{bib6}. Since this paper focuses on transit orbits constructing low-energy transfers, so $H_0$ does not need to be very high. Therefore, we investigate a specific range of $H_0$ at each ${\theta _{{\text{S0}}}}$. The selection criteria are as follows: when $H_0$ is lower than the minimum value (${H_{01}}$) of the selected range, the variation in classification is too dramatic to summarize the characteristics; when $H_0$ is higher than the maximum value (${H_{02}}$) of the selected range, the classification is similar to that under ${H_{02}}$. Consequently, the range is selected from $-847.5100$ to $-847.4945\text{ } {\left( {{\text{LU/TU}}} \right)^2}$ for each investigated ${\theta _{{\text{S0}}}}$ in this paper. First, a global map of classifications with $H_0$ at ${\theta _{{\text{S0}}}} = 45{\text{ }}\deg $ is presented.

\subsubsection{Global map of classification}\label{subsubsec4.1.1}
\begin{figure}[!htb]%
\centering
\includegraphics[width=0.85\textwidth]{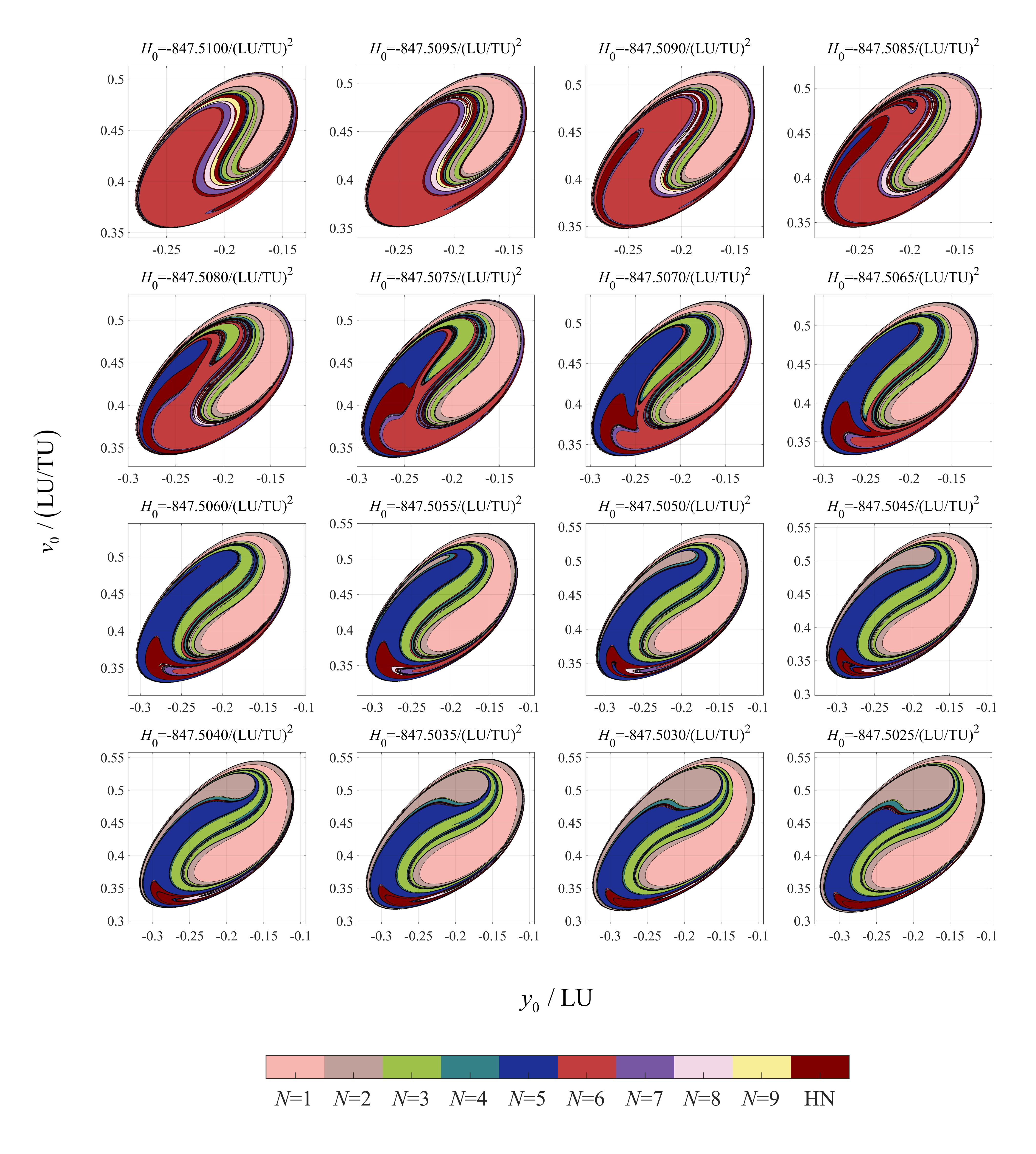}
\caption{Global map of classifications.}\label{fig6}
\end{figure}

\begin{figure}[!htb]%
\centering
\includegraphics[width=0.85\textwidth]{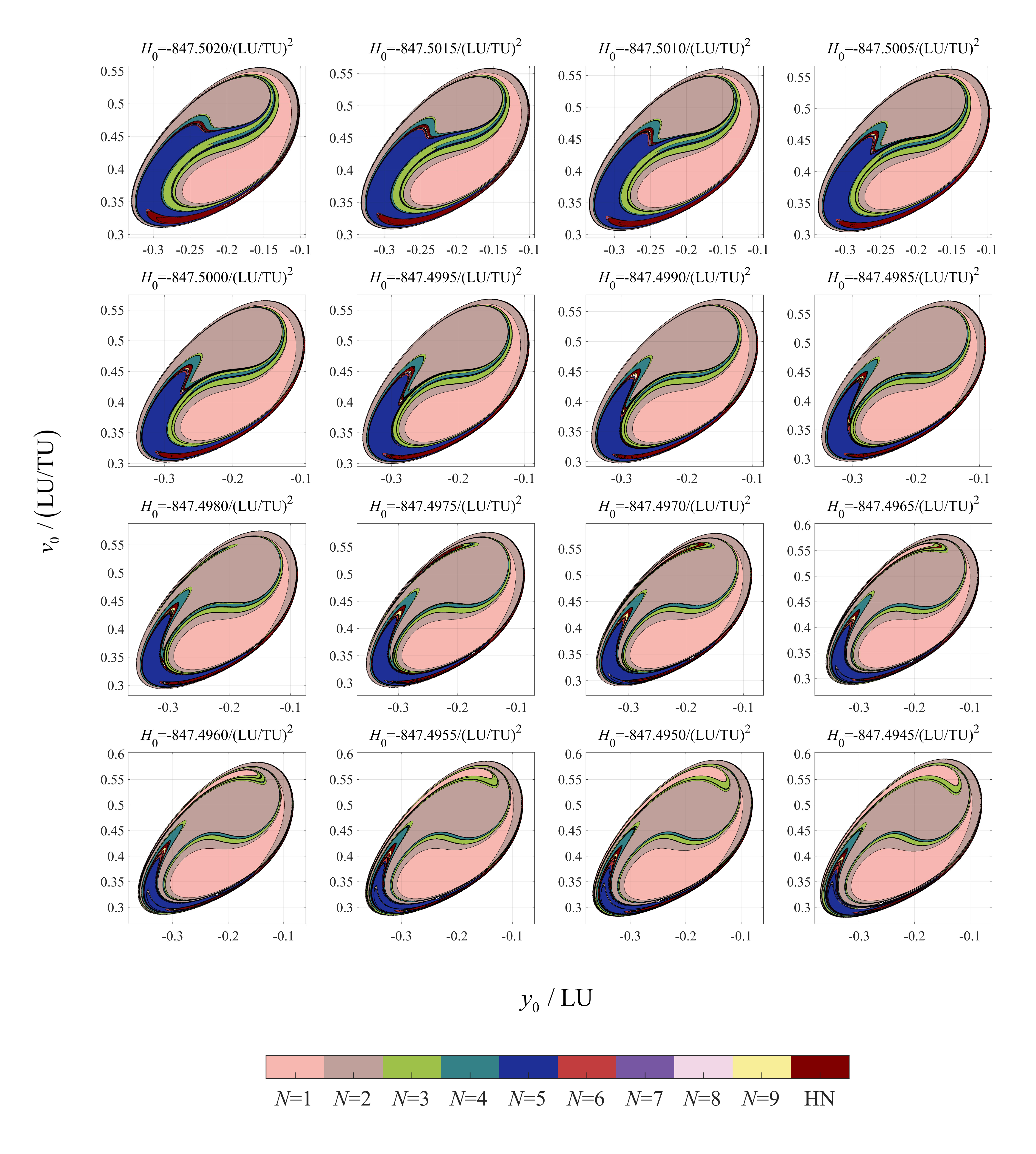}
\caption{Global map of classifications.}\label{fig7}
\end{figure}

Figures \ref{fig6} and \ref{fig7} show the global map of classifications under different $H_0$ at ${\theta _{{\text{S0}}}} = 45{\text{ }}\deg $.  The regions with different colors show the different \textit{N}, along with the distribution of different transit orbit families inside the LCSs. In addition, regions highlighted in dark red are defined as high-\textit{N} (HN) regions (i.e., the regions with $N>10$). Note that the HN regions are usually associated with local scatter distribution, which likely corresponds to chaotic scattering and the theory of leaking Hamiltonian systems \citep{bib15}. Moreover, the transit orbits related to HN regions are usually associated with the lunar collision trajectories (i.e., the states \textbf{\textit{X}} satisfy $\sqrt {{{\left( {x - \mu  + 1} \right)}^2} + {y^2}}  < {R_{\text{M}}}$) \citep{bib23}. When the transit orbits are used to construct low-energy transfers, the lunar collision trajectories should be excluded. Therefore, this paper focuses on other typical families of transit orbits.

Other families of transit orbits are generally separated by the HN regions and are collectively referred to as regular regions. In the regular regions, different \textit{N}  values indicate different families of transit orbits. Furthermore, the families with the same \textit{N} may not be distributed continuously and can be considered as different families. The geometry and characteristics of transit orbits differ significantly between different families. 

As shown in Figs. \ref{fig6} and \ref{fig7}, when $H_0$ varies from $-847.5095$ to $-847.5090 {\text{ }} {\left( {{\text{LU/TU}}} \right)^2}$, an island of HN region emerges from the family with $N=6$ and gradually expands into a larger HN region (i.e., an ocean of HN region). When $H_0$ varies from $-847.5090$ to $-847.5080 {\text{ }} {\left( {{\text{LU/TU}}} \right)^2}$, a family island with $N=5$ (blue region) emerges from the ocean of HN region. Meanwhile, a family with \textit{N}=3 (green region) begins to appear, which is associated with L2 escape trajectories, detailed in Subsections \ref{subsubsec4.1.3}-\ref{subsubsec4.1.6}. As the values of $H_0$ increase (from $-847.5085$ to $-847.5065 {\text{ }} {\left( {{\text{LU/TU}}} \right)^2}$ ), the family island with $N=5$ (blue region) gradually expands to a family ocean. Meanwhile, the area of the family with $N=6$ (red region) shrinks gradually and disappears. At the same time, the family with $N=3$ (green region) also expands gradually and occupies the central regions of the LCS. After the area of the aforementioned family with $N=5$  occupies the lower left regions of the LCS and continues to expand with the increase of $H_0$, new islands of HN regions are generated. These new islands eventually form families with $N=2$ and $N=1$, which are associated with L2 escape trajectories (see details in Subsections \ref{subsubsec4.1.3}-\ref{subsubsec4.1.6}).
Based on the global map of classifications at ${\theta _{{\text{S0}}}} = 45{\text{ }}\deg $, four typical classifications under four different levels of $H_0$ at ${\theta _{{\text{S0}}}} = 45{\text{ }}\deg $ are analyzed in detail. The classifications under other $H_0$ are similar to these four typical classifications. The four different levels of $H_0$ are set to $ -847.5100,{\text{ }}-847.5075,{\text{ }}-847.5025,{\text{ }}\text{and} -847.4950{\text{ }}{({\text{LU/TU}})^2}$ (the four $H_0$ values are denoted as IH I, IH II, IH III, and IH IV for the captions of subsections). Under these four levels of $H_0$, initial state sets of transit orbit families are extracted, and corresponding transfer characteristics are analyzed.
\subsubsection{Classifications under IH I}\label{subsubsec4.1.2}
\begin{figure}[!htb]%
\centering
\includegraphics[width=0.8\textwidth]{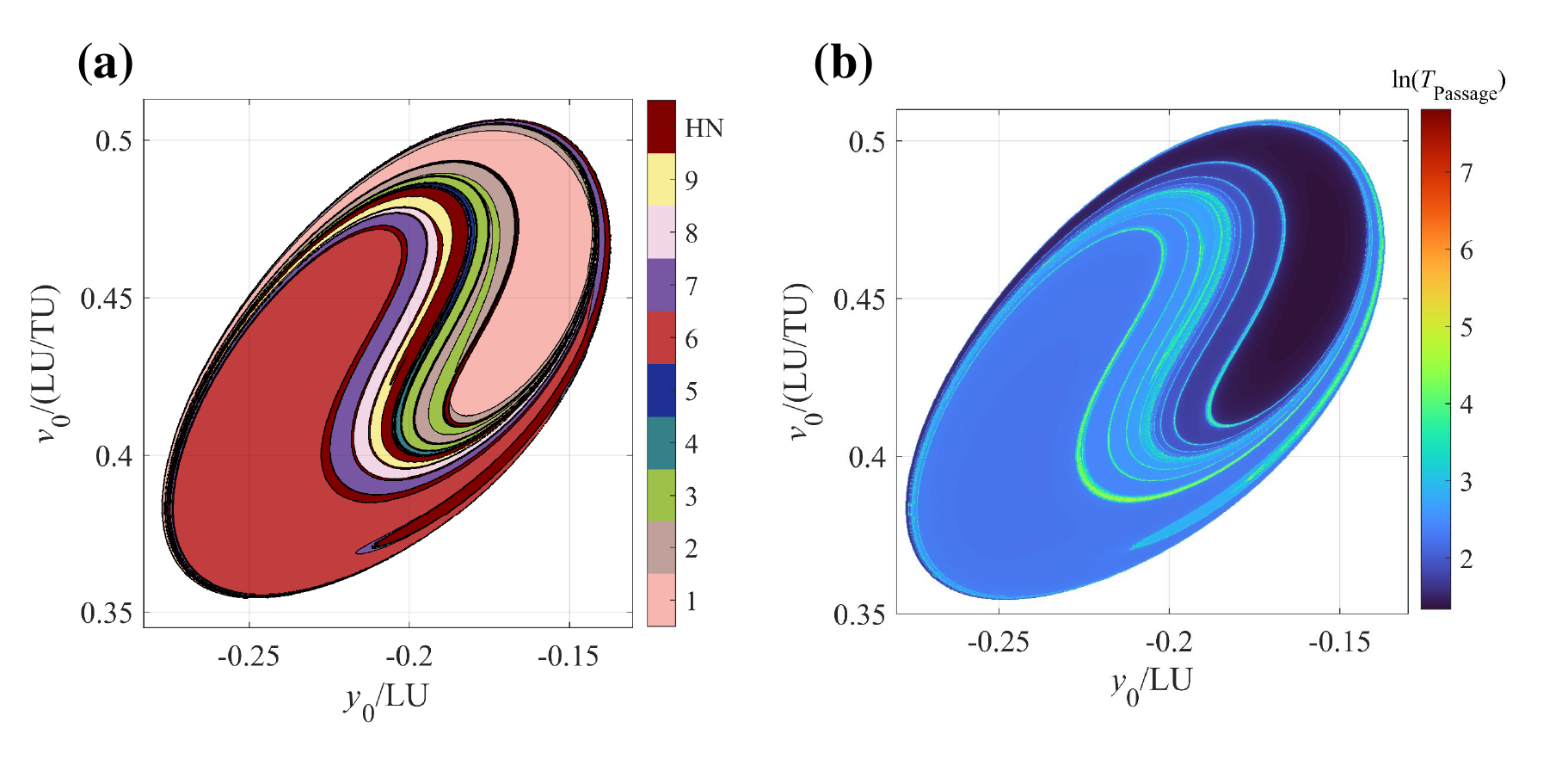}
\caption{Consistency between two classification methods. (a) Classification results based on \textit{N}; (b) Classification results based on ${T_{{\text{Passage}}}}$.}\label{fig8}
\end{figure}

\begin{figure}[!htb]%
\centering
\includegraphics[width=0.4\textwidth]{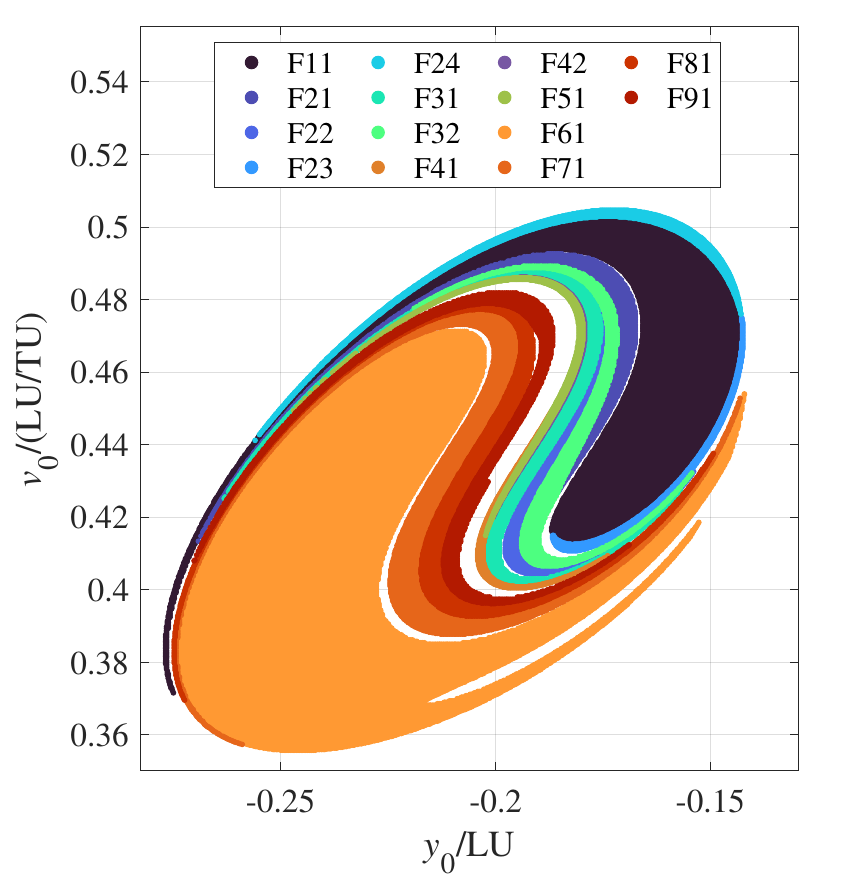}
\caption{Initial state sets of transit orbit families.}\label{fig9}
\end{figure}

Figure \ref{fig8} (a) presents the classification under ${H_0} =  - 847.5100{\text{ }}{({\text{LU/TU}})^2}$. The classification results are consistent with those based on the ${T_{{\text{Passage}}}}$ \citep{bib15,bib34}, as shown in Fig. \ref{fig8} (b). However, the classification results based on \textit{N} exhibit clear boundaries between different families. Subsequently, based on the classification results shown in Fig. \ref{fig8} (a), the initial state sets of transit orbit families in the regular regions are extracted and presented in Fig. \ref{fig9}, where families are marked as F11, F12, and so on. The first number denotes the \textit{N} value, while the second number distinguishes different families with the same \textit{N}.

\begin{figure}[!htb]%
\centering
\includegraphics[width=0.9\textwidth]{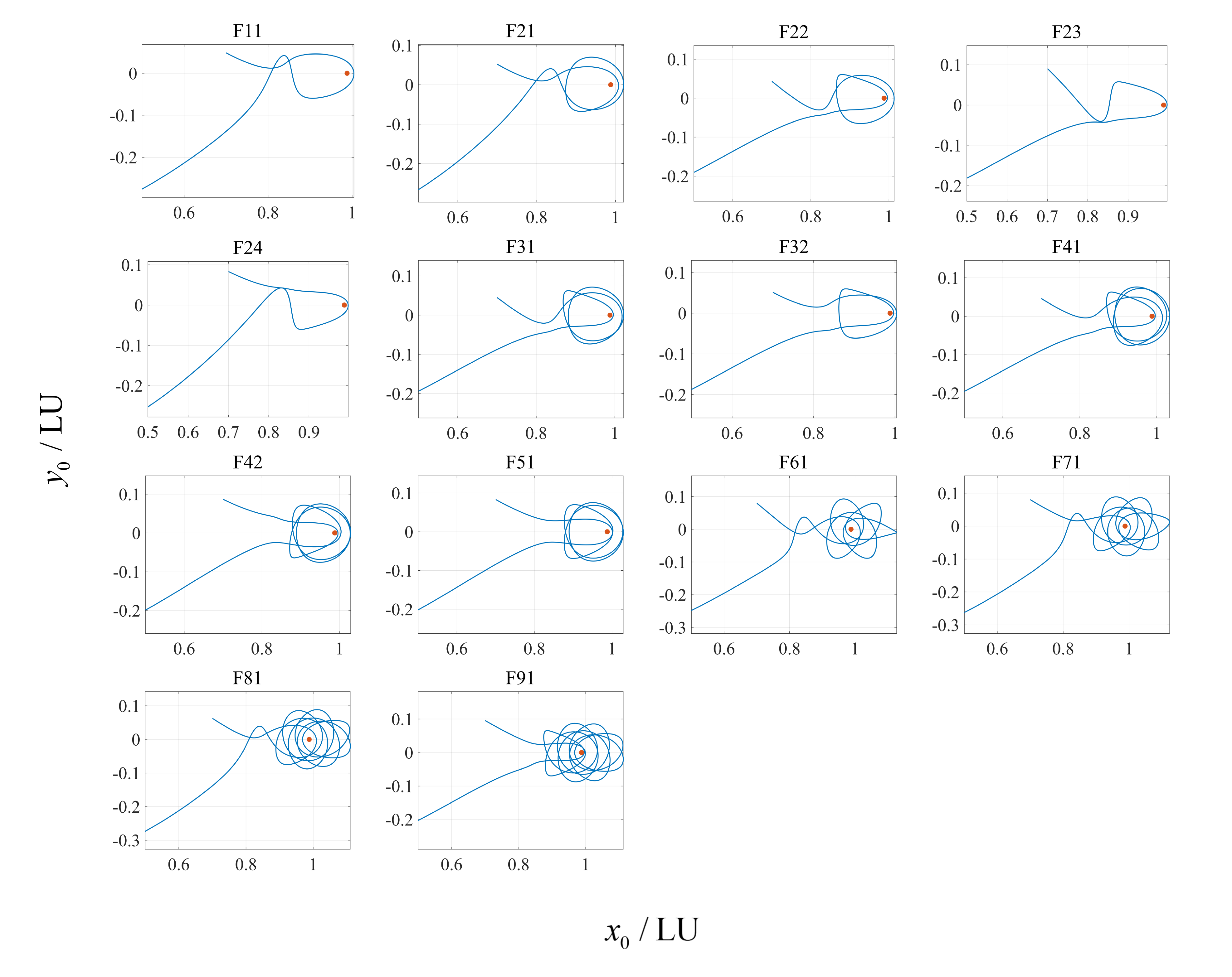}
\caption{Typical patterns of transit orbits associated with different families.}\label{fig10}
\end{figure}

Figure \ref{fig9} shows 14 families in the regular regions. The typical patterns of transit orbits associated with these 14 families are shown in Fig. \ref{fig10}. All the transit orbits associated with regular regions are L1 escape trajectories because the low $H_0$ makes the L2 region fail to open. It is found that transit orbits associated with families F23 and F24 are similar to those associated with F11 in terms of patterns. This similarity arises because family F11 is in a bordering position with the other two families, and the continuous dependence on initial conditions of solutions to ordinary differential equations \citep{bib46} results in similar dynamical behaviors of the transit orbits. Additionally, transit orbits associated with families F11, F21, F22, F23, F24, F31, F32, F41, F42, and F51 can be considered as short-term capture trajectories, while families F61, F71, F81, and F91 can be considered as tour trajectories \citep{bib23,bib60}.

\begin{figure}[!htb]%
\centering
\includegraphics[width=0.9\textwidth]{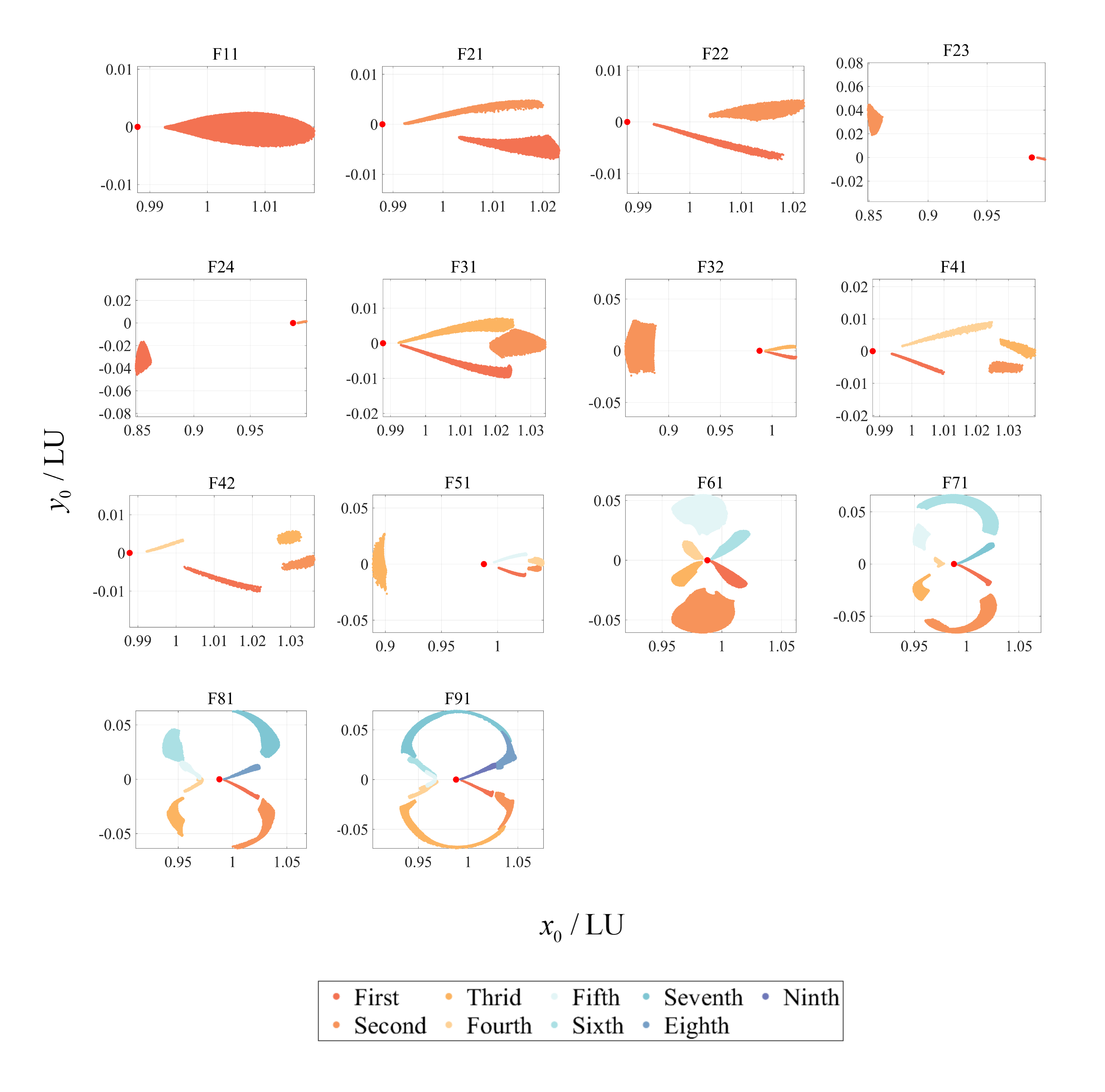}
\caption{Periapsis distribution of transit orbit families.}\label{fig11}
\end{figure}

Figure \ref{fig11} shows the periapsis distribution about the Moon for different families. We find that the periapses of short-term capture trajectories have a wedge-shaped distribution, while the periapses of tour trajectories exhibit a distribution encircling the Moon. Moreover, families F23 and F24 have a similar periapsis distribution (i.e., the first periapses of family F23 and the second periapses of family F24) compared with that of F11. Therefore, transit orbits associated with families F23 and F24 can be considered as distorted trajectories derived from those associated with family F11.
Since $H_0$ is so low that the L2 region fails to open, all the transit orbits are L1 escape trajectories. When $H_0$ is sufficiently high, the L2 region will open and L2 escape trajectories will be generated. Subsequently, the classifications and different dynamical phenomena under higher $H_0$ values are presented.
\subsubsection{Classifications under IH II}\label{subsubsec4.1.3}
\begin{figure}[!htb]%
\centering
\includegraphics[width=0.4\textwidth]{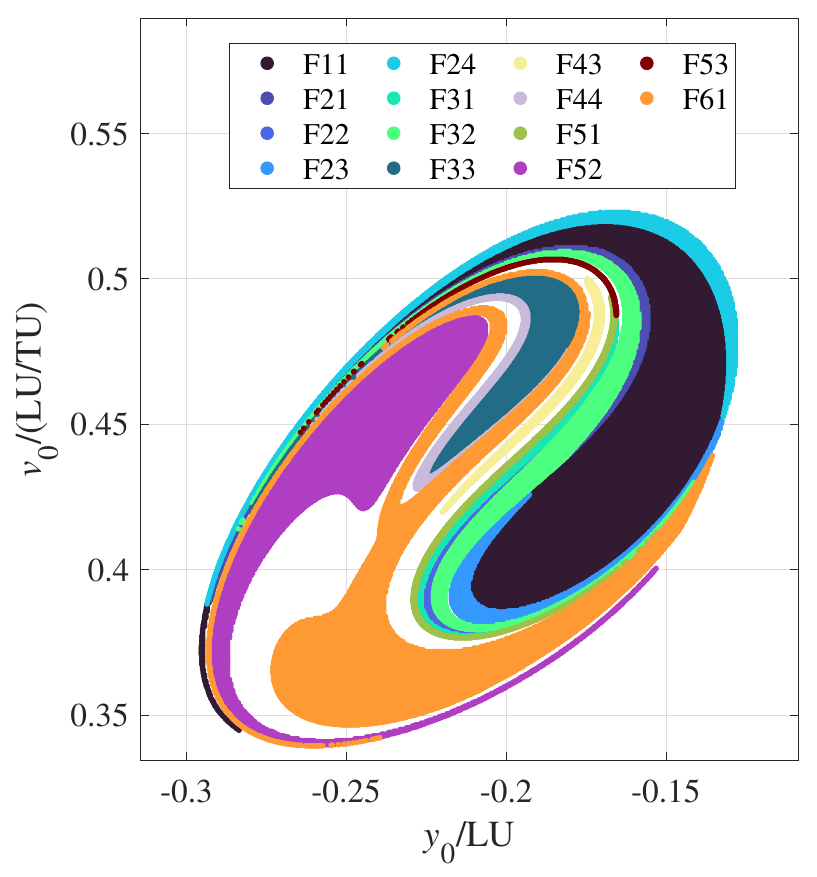}
\caption{Initial state sets of transit orbit families.}\label{fig12}
\end{figure}

\begin{figure}[!htb]%
\centering
\includegraphics[width=0.9\textwidth]{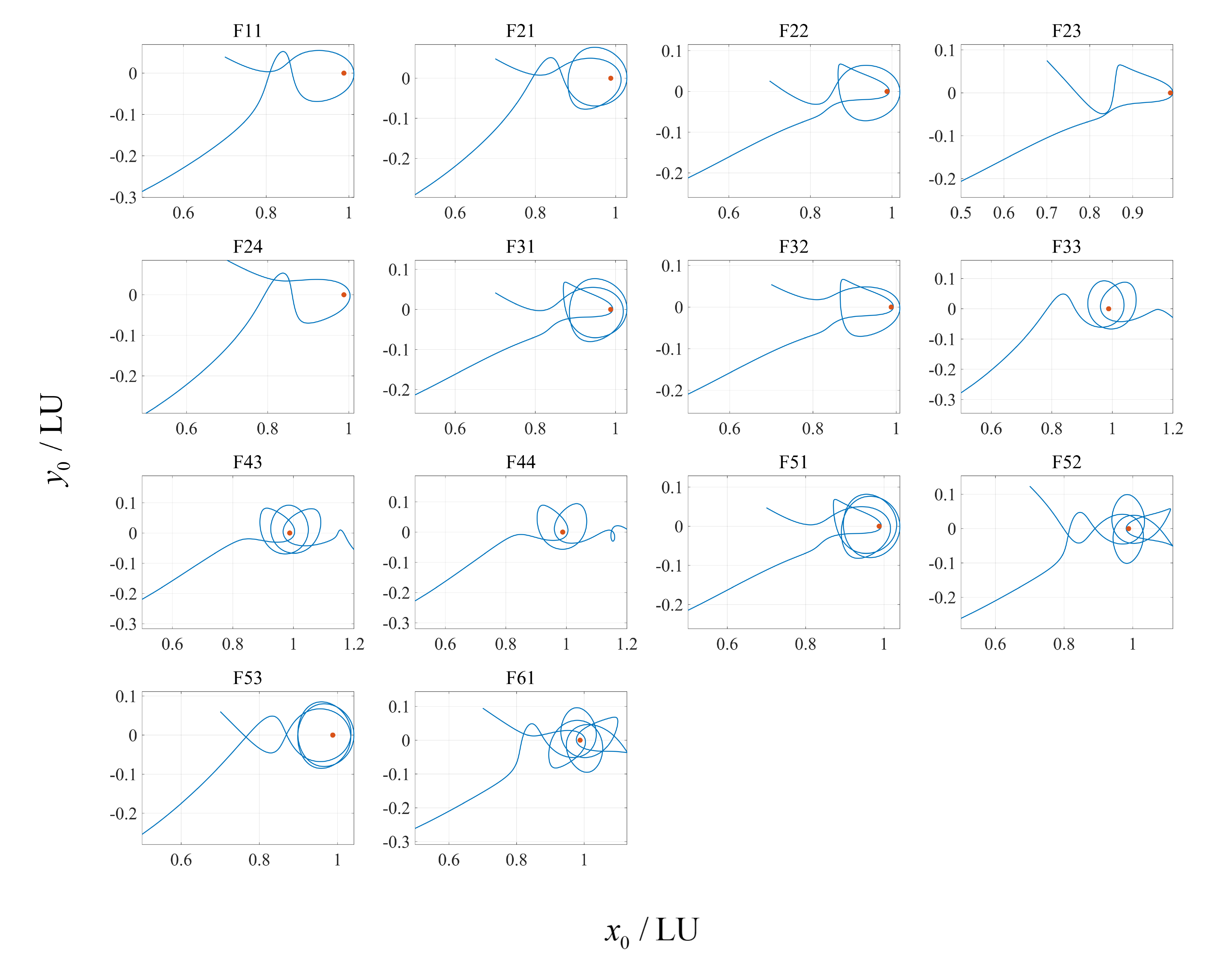}
\caption{Typical patterns of transit orbits associated with different families.}\label{fig13}
\end{figure}

\begin{figure}[!htb]%
\centering
\includegraphics[width=0.9\textwidth]{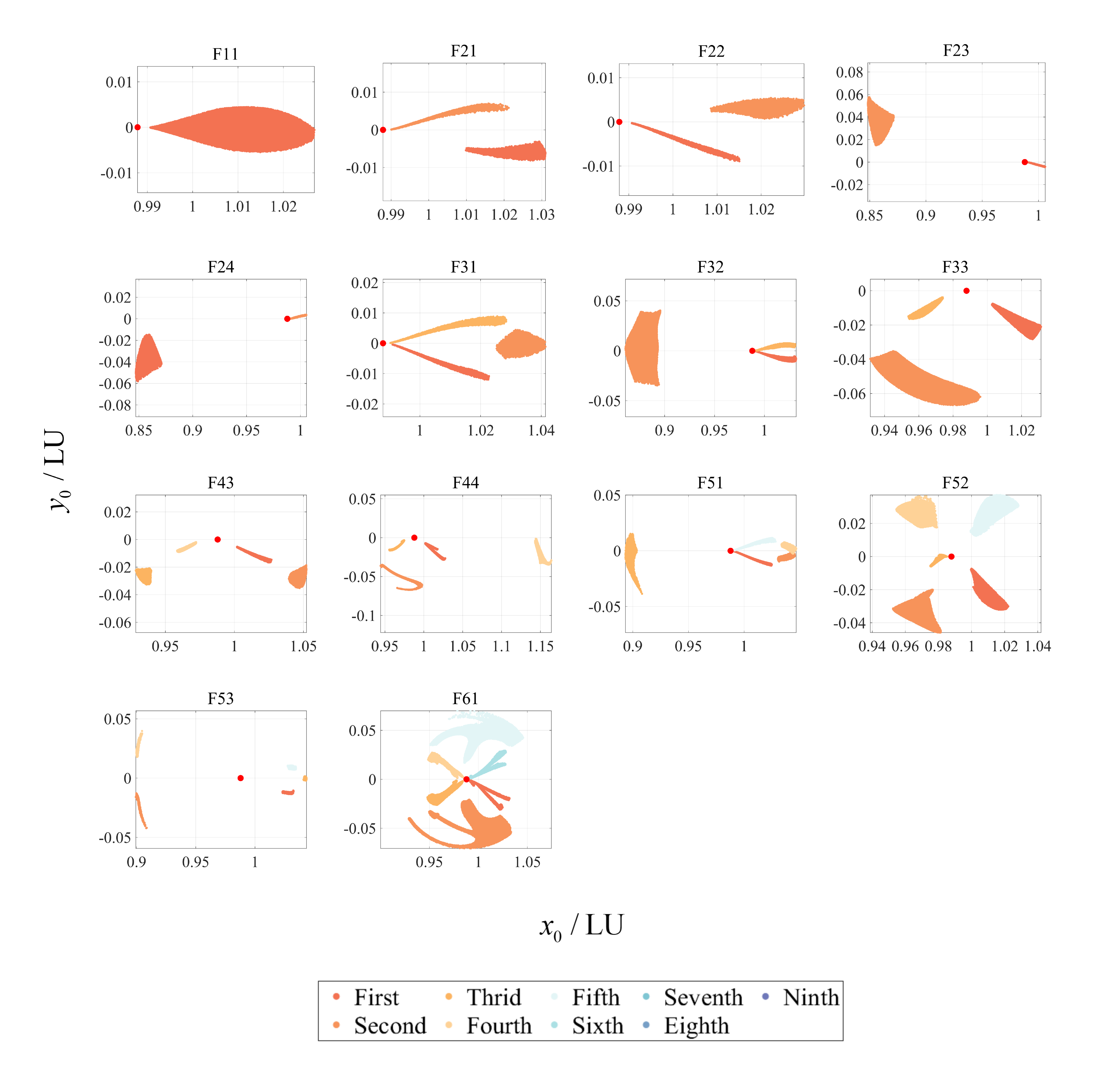}
\caption{Periapsis distribution of transit orbit families.}\label{fig14}
\end{figure}

Figure \ref{fig12} shows 14 families in the regular regions. For L1 escape trajectories, short-term capture trajectories are generated in families F11, F21, F22, F23, F24, F31, F32, F51, and F53, while tour trajectories are generated in families F52 and F61, as shown in Figs. \ref{fig13} and \ref{fig14}.
As shown in Fig. \ref{fig13}, it is found that L2 escape trajectories are generated in families F33, F43, and F44. Most of the periapses of L2 escape trajectories satisfy $y_p < 0$  (see Fig. \ref{fig14}). It is concluded that the dynamical behaviors of transit orbits can be clearly revealed by the classifications based on $N$ and periapsis distribution.

\subsubsection{Classifications under IH III}\label{subsubsec4.1.4}
\begin{figure}[!htb]%
\centering
\includegraphics[width=0.4\textwidth]{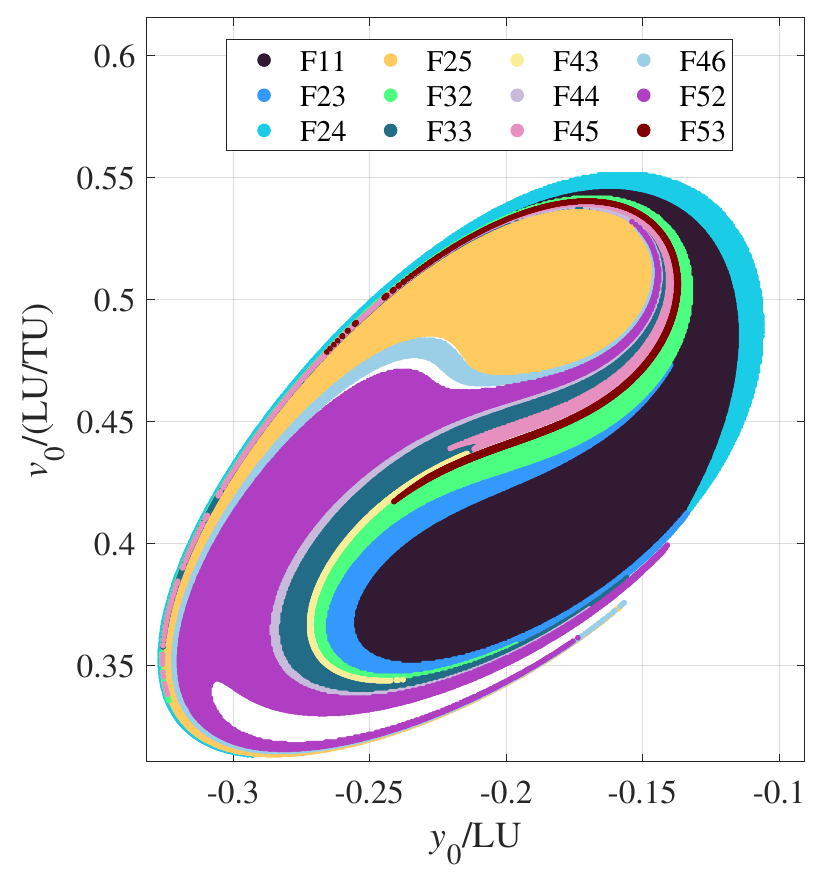}
\caption{Initial state sets of transit orbit families.}\label{fig15}
\end{figure}

\begin{figure}[!htb]%
\centering
\includegraphics[width=0.9\textwidth]{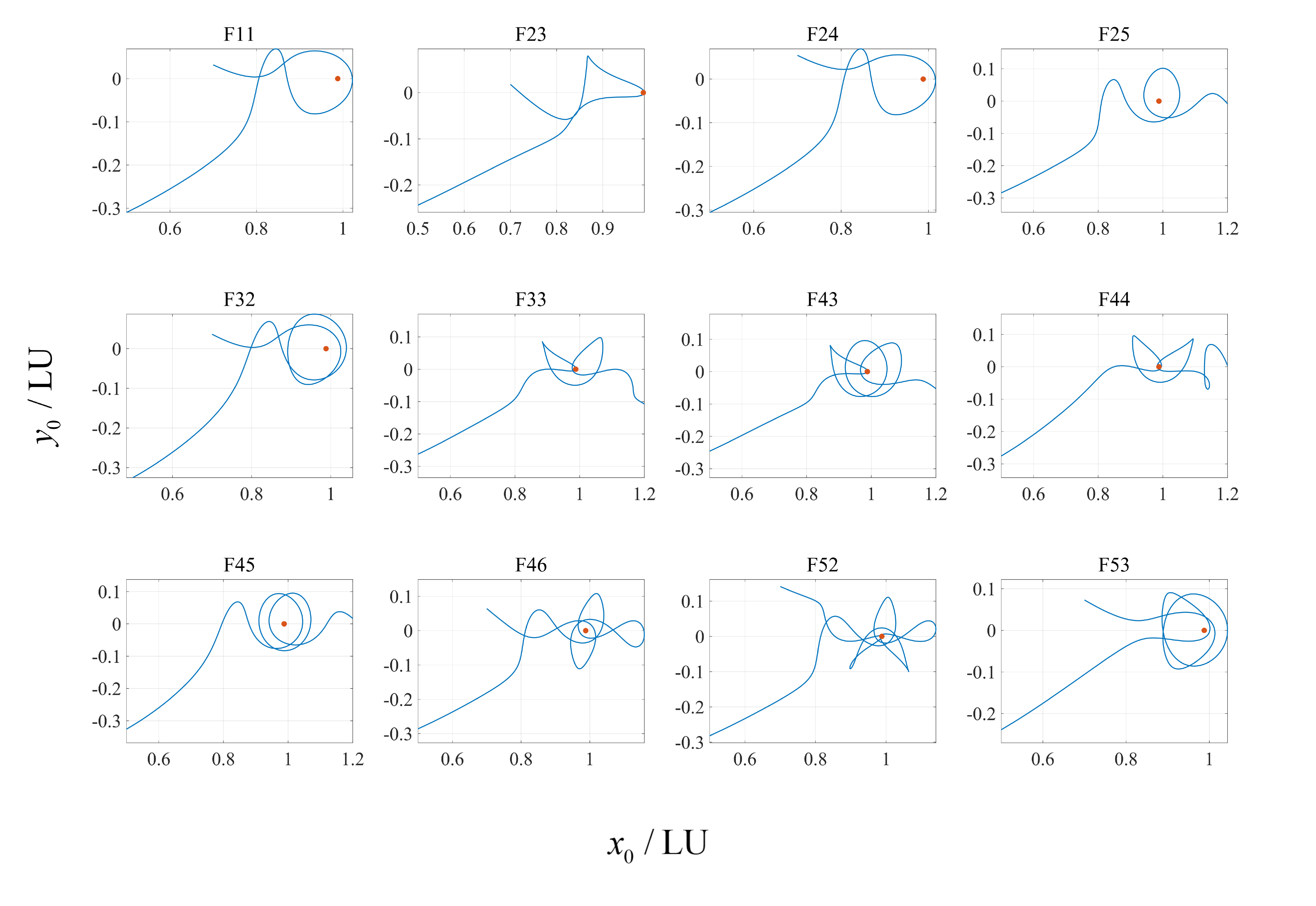}
\caption{Typical patterns of transit orbits associated with different families.}\label{fig16}
\end{figure}

\begin{figure}[!htb]%
\centering
\includegraphics[width=0.9\textwidth]{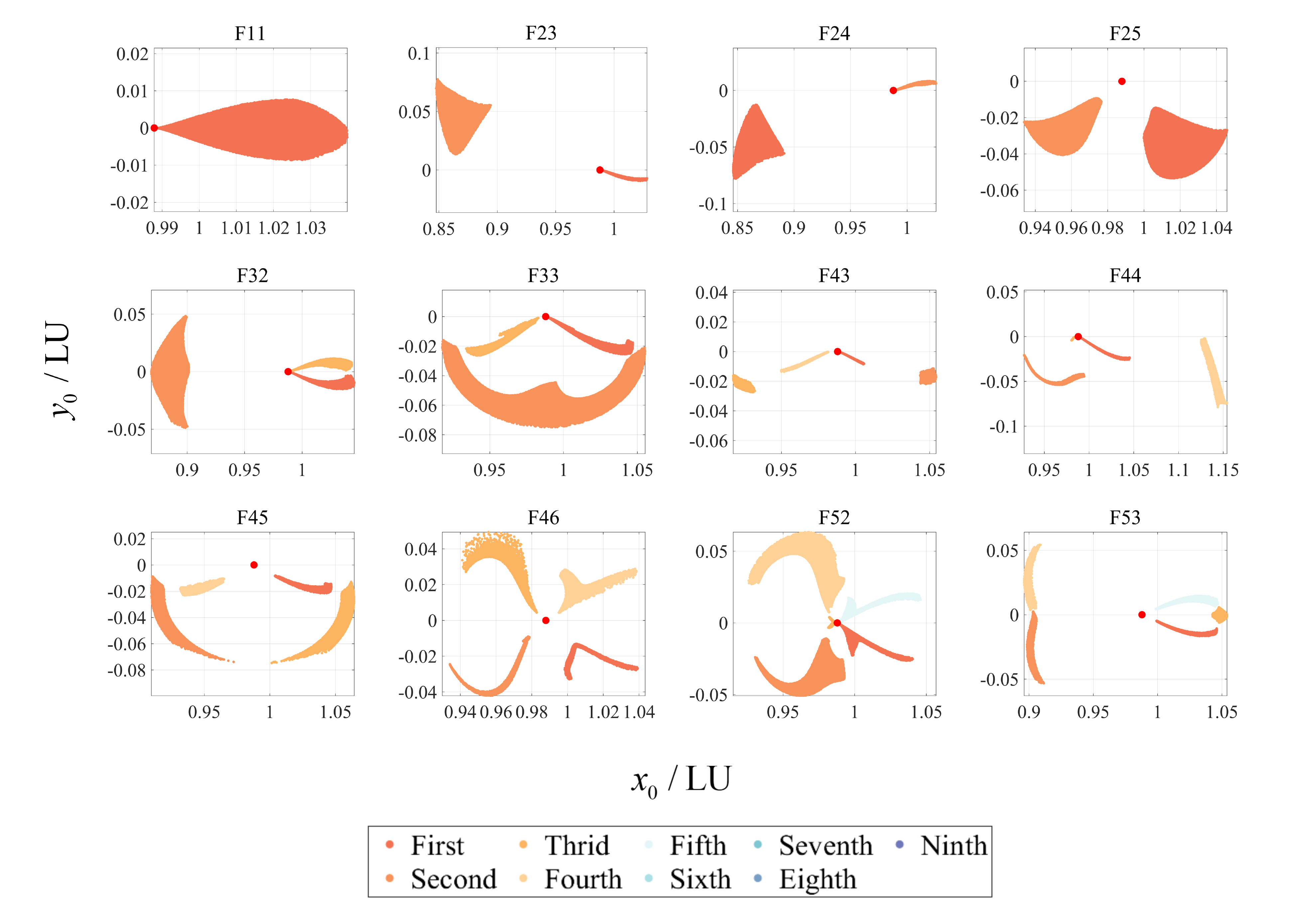}
\caption{Periapsis distribution of transit orbit families.}\label{fig17}
\end{figure}

When $H_0$ continues to increase, as shown in Fig. \ref{fig15}, there are 12 families in the regular regions. Figure \ref{fig16} shows typical trajectory patterns. For L1 escape trajectories, short-term capture trajectories are generated in families F11, F23, F24, F32, F51, and F53, while tour trajectories are generated in families F46 and F52. The L2 escape trajectories are generated in families F25, F33, F43, F44, and F45. With the increase of $H_0$, there is an overall decreasing trend in \textit{N}, which is reflected in the shrinkage of families of relatively high \textit{N}. Additionally, the percentages of L2 escape trajectories increase when $H_0$ increases, while percentages and \textit{N} of tour trajectories decrease. Figure \ref{fig17} presents the periapsis distribution of the transit orbit families.


\subsubsection{Classifications under IH IV}\label{subsubsec4.1.5}
\begin{figure}[!htb]%
\centering
\includegraphics[width=0.4\textwidth]{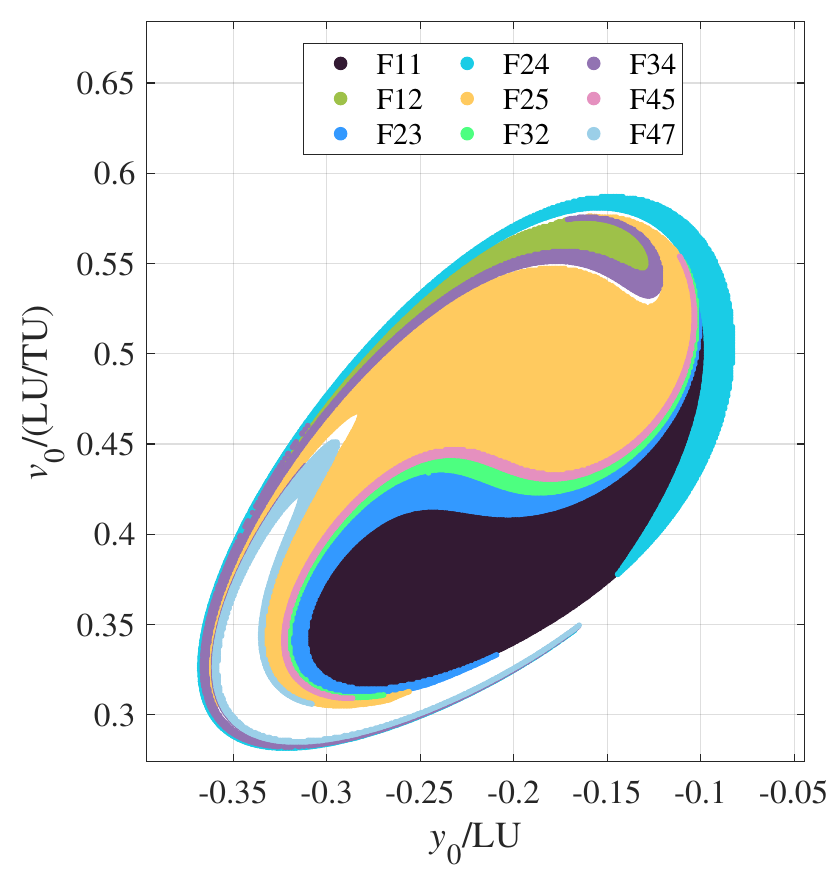}
\caption{Initial state sets of transit orbit families.}\label{fig18}
\end{figure}

\begin{figure}[!htb]%
\centering
\includegraphics[width=0.9\textwidth]{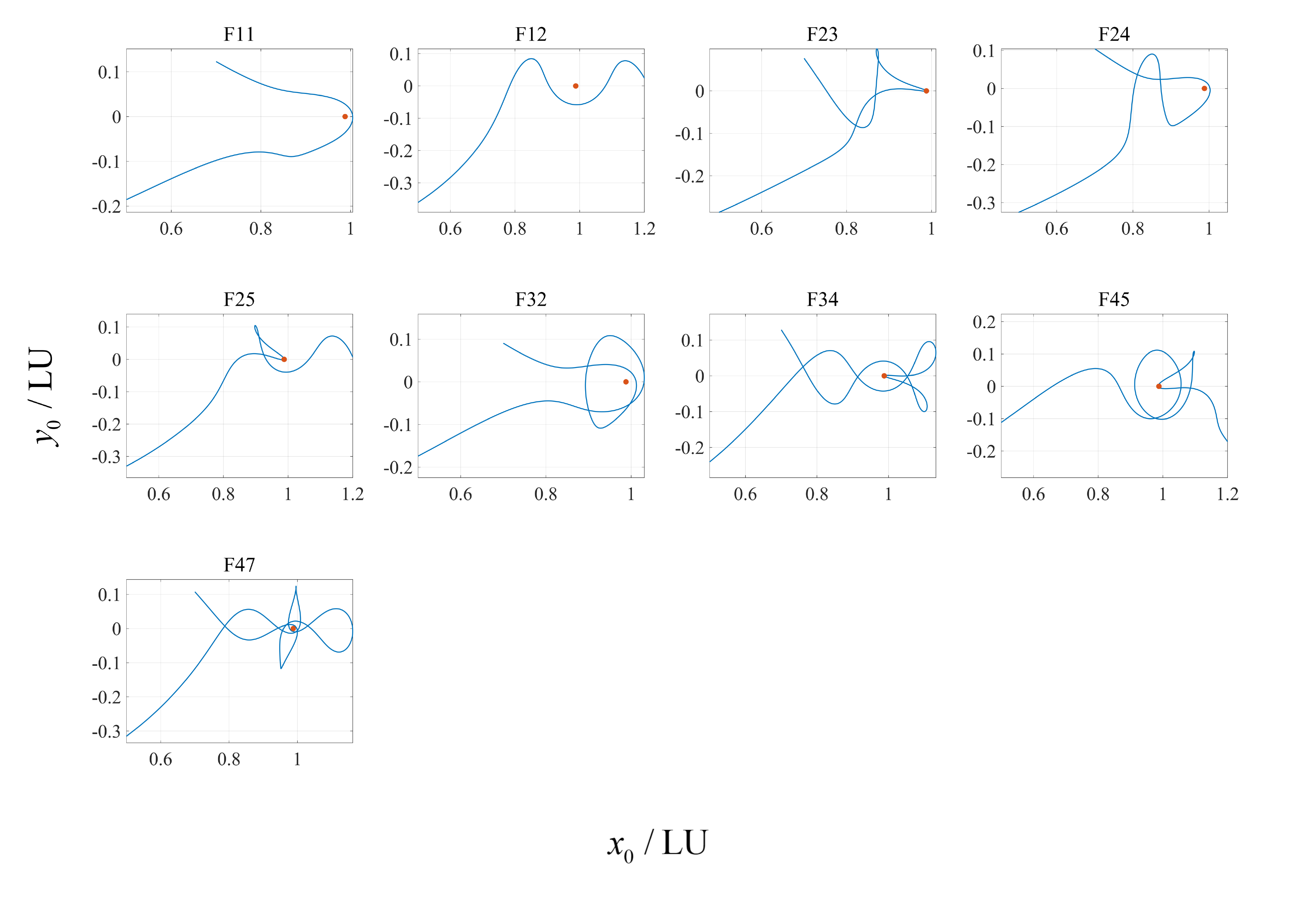}
\caption{Typical patterns of transit orbits associated with different families.}\label{fig19}
\end{figure}

\begin{figure}[!htb]%
\centering
\includegraphics[width=0.9\textwidth]{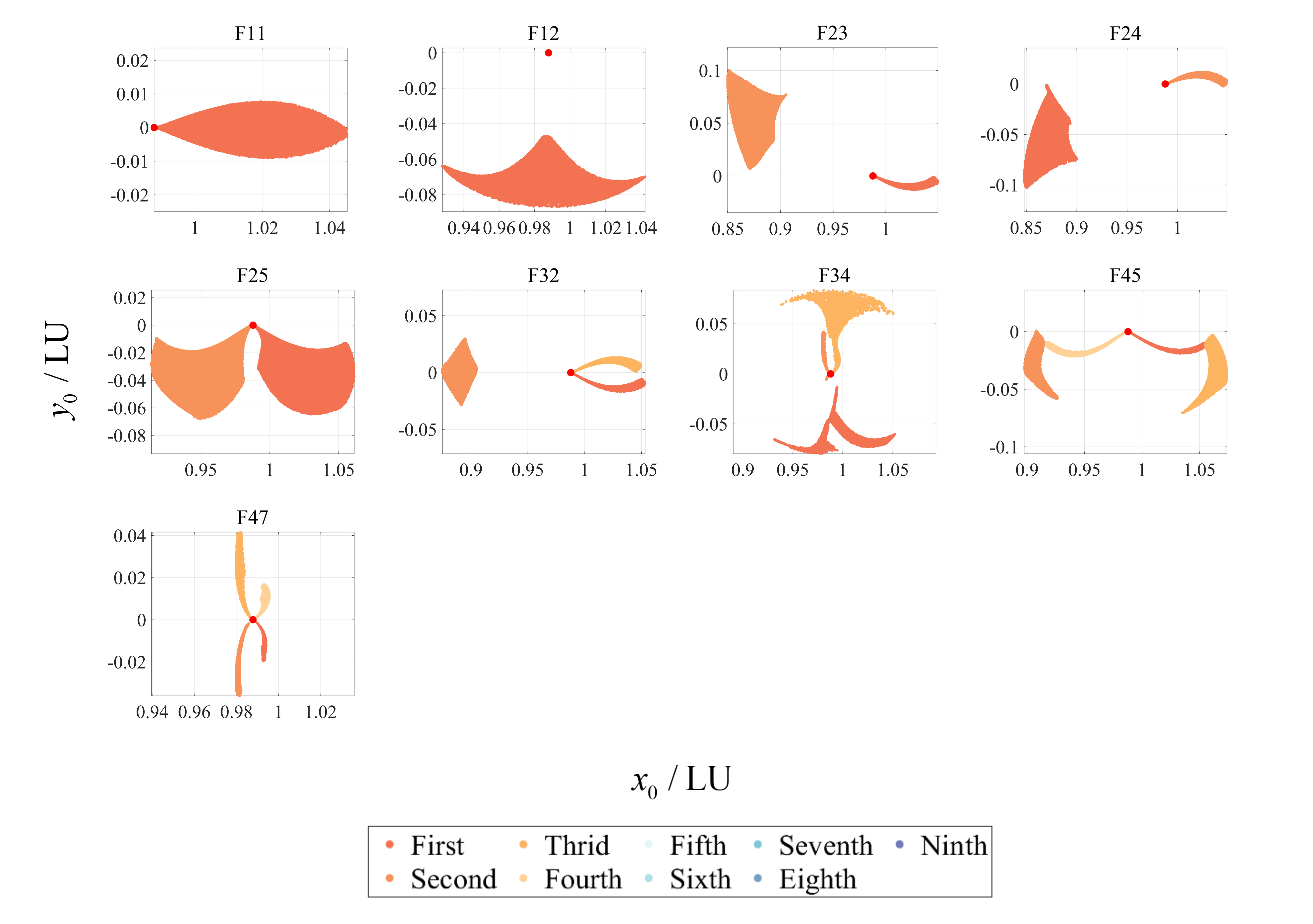}
\caption{Periapsis distribution of transit orbit families.}\label{fig20}
\end{figure}

Figure \ref{fig18} shows 9 families in the regular regions. For L1 escape trajectories, short-term capture trajectories are generated in families F11, F23, F24, and F32, while tour trajectories are generated in families F34 and F47 (see Figs. \ref{fig19} and \ref{fig20}). L2 escape trajectories are generated in families F12, F25, and F34. Linking the classification with those under the three aforementioned $H_0$ values, it is found that variations in $H_0$ are accompanied by the emergence of new families and the disappearance of existing families. In the next subsection, the evolution laws of classifications with respect to $H_0$ are discussed and analyzed.
\subsubsection{Evolution laws of classifications with respect to the initial Hamiltonian}\label{subsubsec4.1.6}
\begin{figure}[!htb]%
\centering
\includegraphics[width=0.9\textwidth]{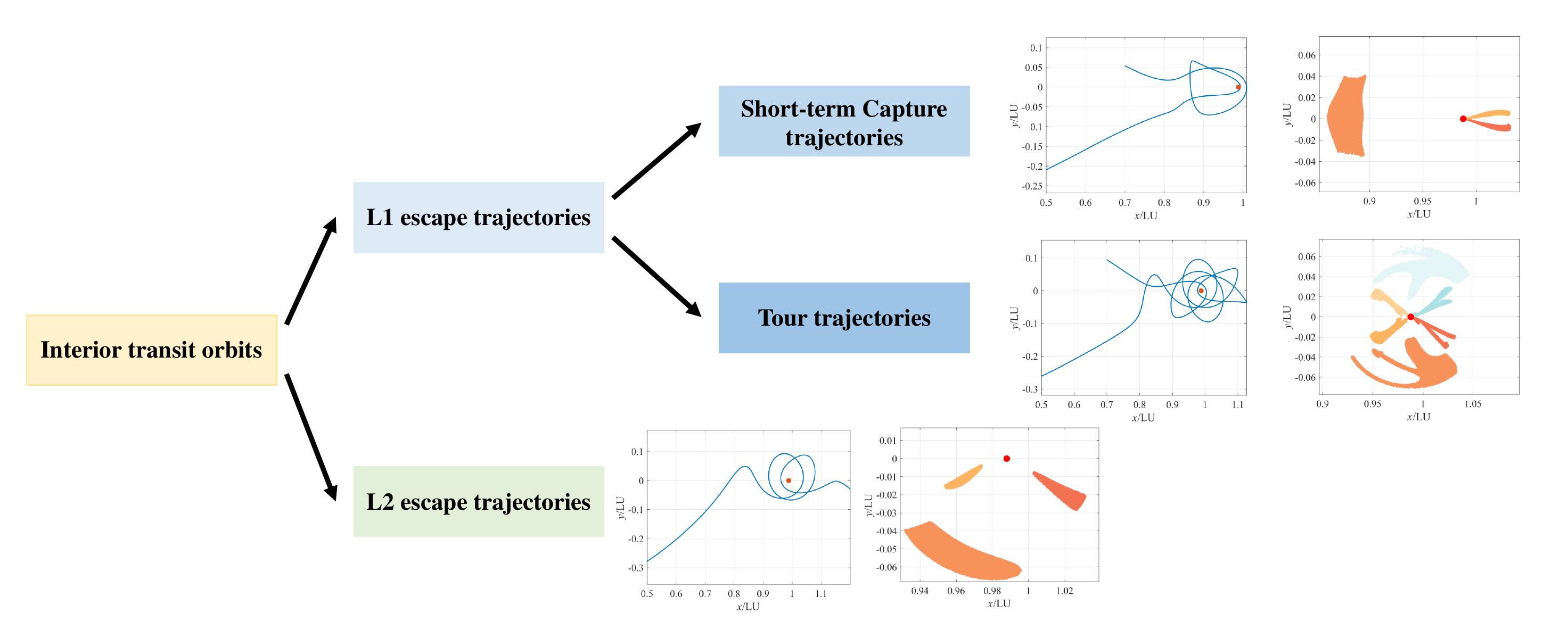}
\caption{Three main types of transit orbits and periapsis distributions.}\label{fig21}
\end{figure}

\begin{figure}[!htb]%
\centering
\includegraphics[width=0.9\textwidth]{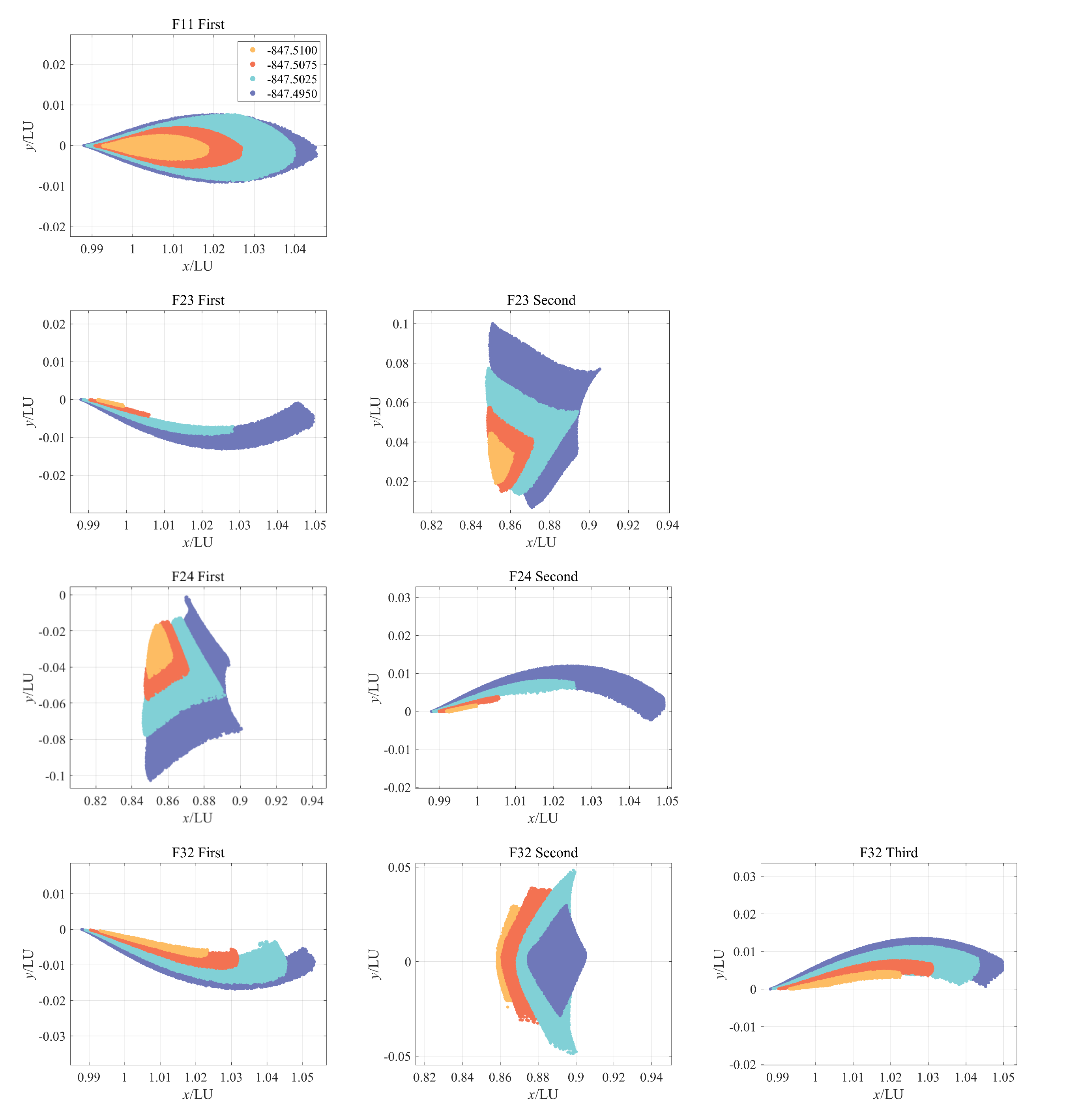}
\caption{The ranges of $h_p$ for four families under four $H_0$ values.}\label{fig22}
\end{figure}

Based on the aforementioned results, we find that transit orbits can be categorized into L1 and L2 escape trajectories according to the termination Poincaré sections (i.e., $U_1$ and $U_2$). Furthermore, L1 escape trajectories can be categorized into short-term capture trajectories and tour trajectories based on the patterns of transit orbits. Therefore, the investigated transit orbits can be categorized into three main types: (i) short-term capture trajectories, (ii) tour trajectories, and (iii) L2 escape trajectories. Under the investigated range of $H_0$, L1 escape trajectories always exist, while the existence of L2 escape trajectories requires a higher $H_0$. The typical trajectory patterns and periapsis distributions are shown in Fig. \ref{fig21}. Subsequently, the evolution laws of $H_0$ are discussed and analyzed in terms of these three types.
For short-term capture trajectories, it is found that families F11, F23, F24, and F32 exist across all the investigated $H_0$ values. Therefore, the percentages, $\left\langle {{T_{{\text{Passage}}}}} \right\rangle $, and the ranges of $h_p$ under these four $H_0$ values are presented to reveal the effects of $H_0$. As shown in Table \ref{tab3}, with the increase of $H_0$, the percentages of families F11 and F32 exhibit an initial increase followed by a decrease, while families F23 and F24 exhibit a continuous increase. Moreover, $\left\langle {{T_{{\text{Passage}}}}} \right\rangle $ of all four families decreases. Figure \ref{fig22} presents the variation of the ranges of $h_p$ of these four families. It is found that, except for the second periapses of family F32, the ranges of $h_p$ increase with the increase of $H_0$. For the second periapsis of family F32, the range of $h_p$ exhibits an initial increase followed by a decrease. Furthermore, the variation of $H_0$ is accompanied by the disappearance of certain short-term capture trajectory families (e.g., families F41, F51, etc.).

\begin{table}[h]
\renewcommand{\arraystretch}{1.5}
\caption{Transfer characteristics of families varying with $H_0$}\label{tab3}%
	\centering

	\begin{tabular}{@{}llll@{}}
 \hline
		
		Family               & ${H_0}/{\left( {{\text{LU/TU}}} \right)^2}$                              & $\eta /\% $    & $\left\langle {{T_{{\text{Passage}}}}} \right\rangle /{\text{TU}}$ \\ \hline
		\multirow{4}{*}{F11} & \multicolumn{1}{l}{$-847.5100$} & 17.15 & 4.1065   \\
		                     & \multicolumn{1}{l}{$-847.5075$} & 22.81 & 3.7921   \\
		                     & \multicolumn{1}{l}{$-847.5025$} & 26.65 & 3.4767   \\
		                     & \multicolumn{1}{l}{$-847.4950$} & 25.48 & 3.2484   \\ \hline
		\multirow{4}{*}{F23} & \multicolumn{1}{l}{$-847.5100$} & 0.50  & 4.6511   \\
		                     & \multicolumn{1}{l}{$-847.5075$} & 1.55  & 4.3471   \\
		                     & \multicolumn{1}{l}{$-847.5025$} & 3.66  & 3.9871   \\
		                     & \multicolumn{1}{l}{$-847.4950$} & 5.85  & 3.7631   \\ \hline
		\multirow{4}{*}{F24} & \multicolumn{1}{l}{$-847.5100$} & 1.05  & 4.5209   \\
		                     & \multicolumn{1}{l}{$-847.5075$} & 2.24  & 4.2252   \\
		                     & \multicolumn{1}{l}{$-847.5025$} & 4.07  & 3.8871   \\
		                     & \multicolumn{1}{l}{$-847.4950$} & 5.99  & 3.7155   \\ \hline
		\multirow{4}{*}{F32} & \multicolumn{1}{l}{$-847.5100$} & 2.17  & 5.7073   \\
		                     & \multicolumn{1}{l}{$-847.5075$} & 4.10  & 5.4504   \\
		                     & \multicolumn{1}{l}{$-847.5025$} & 4.55  & 5.2954   \\
		                     & \multicolumn{1}{l}{$-847.4950$} & 2.35  & 5.2364  \\ 
                       \hline

	\end{tabular}
	
\end{table}

For tour trajectories and L2 escape trajectories, the variation in  $H_0$ mainly affects the emergence of new families (i.e., new families with relatively low \textit{N}) and the disappearance of old families (i.e., old families with relatively high \textit{N}). Overall, the \textit{N} values of transit orbit families decrease with the increase of $H_0$.

\subsection{Case II: $\text{135}{\text{ }}\text{deg} $}\label{subsec4.2}
To illustrate the similarities and differences in the classifications at different ${\theta _{{\text{S0}}}}$, we explore two additional cases (i.e., ${\theta _{{\text{S0}}}} = 135{\text{ }}\deg $ and ${\theta _{{\text{S0}}}} = 225{\text{ }}\deg $) to reveal the effects of the solar gravity perturbation on transit orbits. In this subsection, our exploration continues with the case at ${\theta _{{\text{S0}}}} = 135{\text{ }}\deg $.
\subsubsection{Global map of classification}\label{subsubsec4.2.1}
\begin{figure}[!htb]%
\centering
\includegraphics[width=0.85\textwidth]{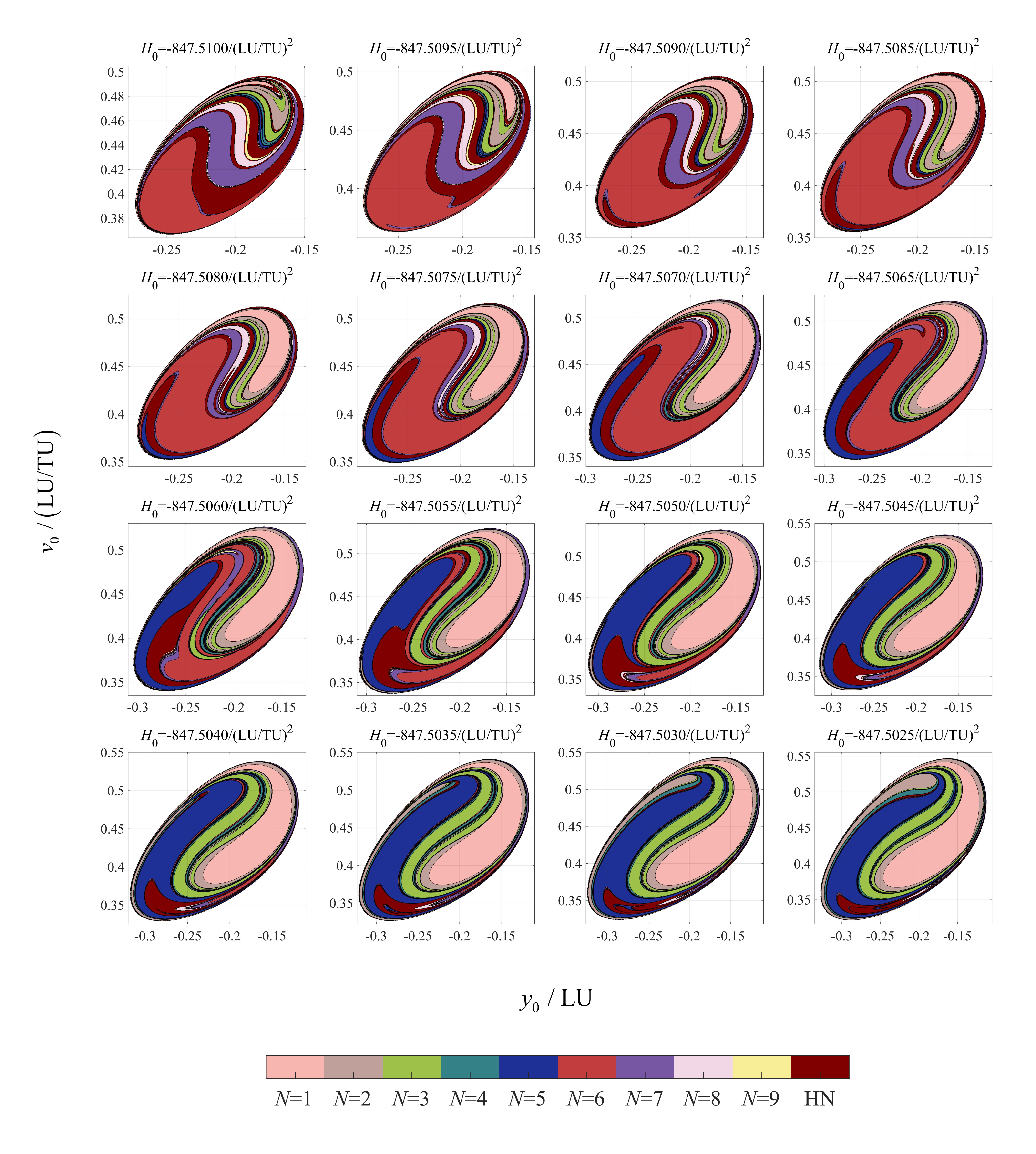}
\caption{Global map of classifications.}\label{fig23}
\end{figure}

\begin{figure}[!htb]%
\centering
\includegraphics[width=0.85\textwidth]{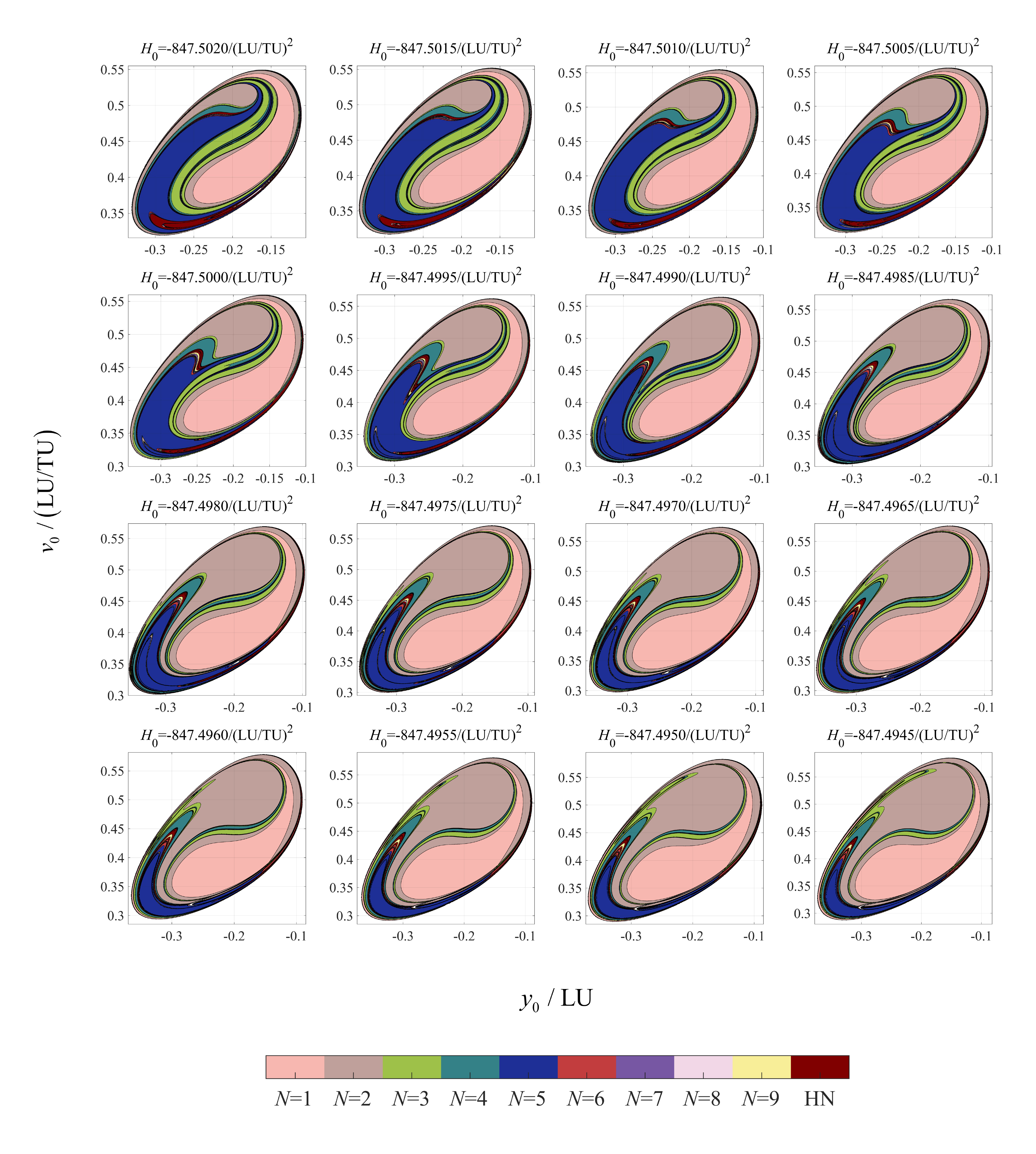}
\caption{Global map of classifications (continued).}\label{fig24}
\end{figure}

Similar to Subsection \ref{subsec4.1}, a global map of classifications with $H_0$ is presented to compare with the cases at ${\theta _{{\text{S0}}}} = 45{\text{ }}\deg $, as shown in Figs. \ref{fig23} and \ref{fig24}. It is found that when $H_0$ is relatively low (e.g. ${H_0} =  - 847.5100{\text{ }}{\left( {{\text{LU/TU}}} \right)^2}$), the difference is that at ${\theta _{{\text{S0}}}} = 135{\text{ }}\deg $, the percentage of family F11 (pink regions) is less than that at ${\theta _{{\text{S0}}}} = 45{\text{ }}\deg $. Additionally, when $H_0$ is relatively high (e.g. ${H_0} =  - 847.4950{\text{ }}{\left( {{\text{LU/TU}}} \right)^2}$), the percentage of family F12 at ${\theta _{{\text{S0}}}} = 135{\text{ }}\deg $ is less than that at ${\theta _{{\text{S0}}}} = 45{\text{ }}\deg $. Referring to the effects of $\theta _{{\text{S0}}}$ on the configurations and distributions of LCSs (i.e., the configurations of the LCSs translate along the $H_0$ axis at different $\theta _{{\text{S0}}}$) shown in Fig. \ref{fig5}, it can be concluded that $\theta _{{\text{S0}}}$ affects the energy conditions of generation and disappearance of the families by affecting the configurations of the LCSs under different $H_0$. Subsequently, classifications under the four aforementioned $H_0$  values are detailed.
\subsubsection{Classifications under IH I-IV}\label{subsubsec4.2.2}
\begin{figure}[h]%
\centering
\includegraphics[width=0.8\textwidth]{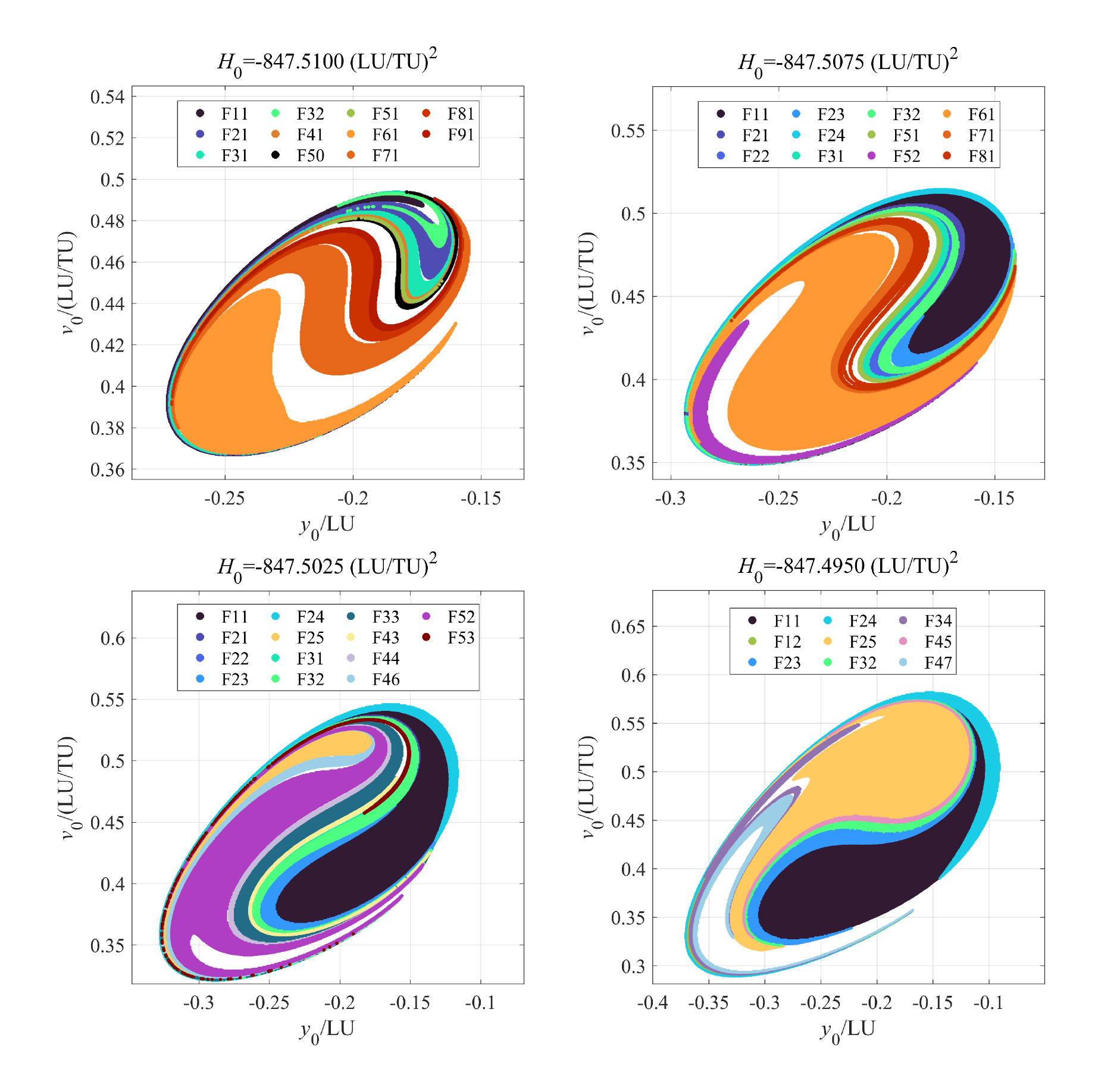}
\caption{Initial state sets of transit orbit families under four $H_0$ values.}\label{fig25}
\end{figure}

\begin{figure}[!htb]%
\centering
\includegraphics[width=0.9\textwidth]{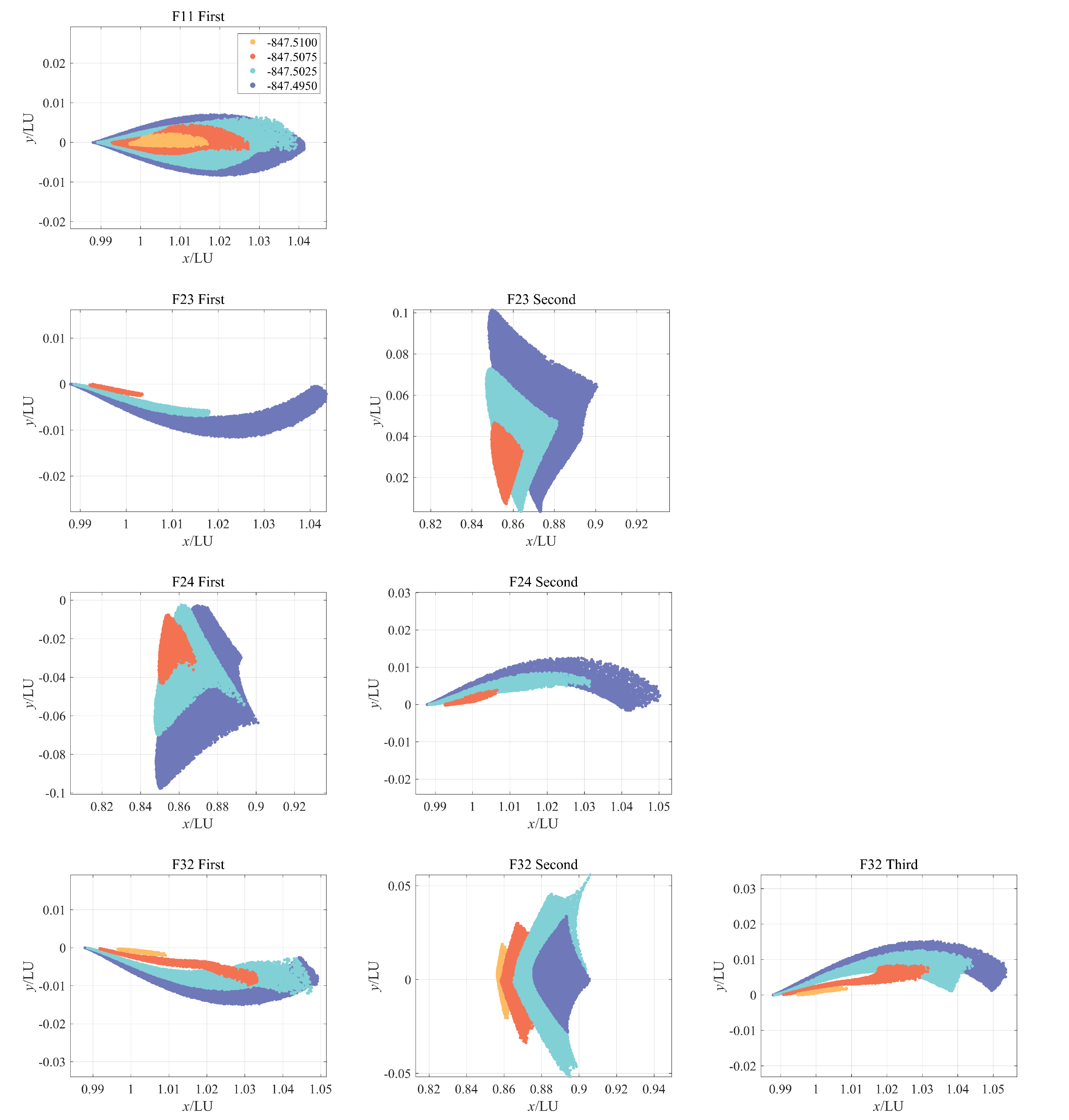}
\caption{The ranges of $h_p$ for four families under four $H_0$ values.}\label{fig26}
\end{figure}

Figure \ref{fig25} presents the extracted initial state sets of classifications under four aforementioned $H_0$ values at ${\theta _{{\text{S0}}}} = 135{\text{ }}\deg $. It is found that the families at ${\theta _{{\text{S0}}}} = 135{\text{ }}\deg $ are similar to those at ${\theta _{{\text{S0}}}} = 45{\text{ }}\deg $, except for family F50. The pattern and periapsis distribution of trajectories associated with family F50 are detailed in Appendix A. Since when $\theta _{{\text{S0}}}$ switches from 45 deg to 135 deg, the configurations of the LCSs translate along the increasing $H_0$ axis (see in Fig. \ref{fig5}), the emergence and disappearance of the same transit orbit families at ${\theta _{{\text{S0}}}} = 135{\text{ }}\deg $ require higher $H_0$  values than at ${\theta _{{\text{S0}}}} = 45{\text{ }}\deg $. For example, families F23 and F24 exist under ${H_0} =  - 847.5100{\text{ }}{({\text{LU/TU}})^2}$ at ${\theta _{{\text{S0}}}} = 45{\text{ }}\deg $, but do not exist under the same $H_0$ at ${\theta _{{\text{S0}}}} = 135{\text{ }}\deg $, requiring higher $H_0$ values, e.g., ${H_0} =  - 847.5075{\text{ }}{({\text{LU/TU}})^2}$ for their existence. 
Table \ref{tab4} presents the percentages and $\left\langle {{T_{{\text{Passage}}}}} \right\rangle $ of typical short-term capture trajectory families (i.e., families F11, F23, F24 and F32) at ${\theta _{{\text{S0}}}} = 135{\text{ }}\deg $. Figure \ref{fig26} shows the ranges of $h_p$ of these aforementioned families. Note that families F23 and F24 do not exist under ${H_0} =  - 847.5100{\text{ }}{({\text{LU/TU}})^2}$. It is found that the trend of parameters varying with $H_0$ for short-term capture trajectories at ${\theta _{{\text{S0}}}} = 135{\text{ }}\deg $ are similar to those at ${\theta _{{\text{S0}}}} = 45{\text{ }}\deg $, except that the percentages of family F11 continuously increase with the increase of $H_0$. In addition, $\left\langle {{T_{{\text{Passage}}}}} \right\rangle $ of these families are longer than those under the same $H_0$ at ${\theta _{{\text{S0}}}} = 45{\text{ }}\deg $. Similarly, the effects of $H_0$ on the families of tour trajectories and L2 escape trajectories are reflected in the emergence and disappearance of transit orbit families. Therefore, it is concluded that the evolution laws with respect to $H_0$ are similar at different $\theta _{{\text{S0}}}$ values but differ in the energy conditions for the emergence and disappearance of transit orbit families.

\begin{table}[h]
\renewcommand{\arraystretch}{1.5}
\caption{Transfer characteristics of families varying with $H_0$}\label{tab4}%
	\centering

	\begin{tabular}{@{}llll@{}}
 \hline
		
		Family               & ${H_0}/{\left( {{\text{LU/TU}}} \right)^2}$                              & $\eta /\% $    & $\left\langle {{T_{{\text{Passage}}}}} \right\rangle /{\text{TU}}$ \\ \hline
		\multirow{4}{*}{F11} & \multicolumn{1}{l}{$-847.5100$} & 1.88 & 5.2976   \\
		                     & \multicolumn{1}{l}{$-847.5075$} & 14.72 & 4.0889   \\
		                     & \multicolumn{1}{l}{$-847.5025$} & 25.72 & 3.5565   \\
		                     & \multicolumn{1}{l}{$-847.4950$} & 27.56 & 3.2676   \\ \hline
		\multirow{4}{*}{F23} & \multicolumn{1}{l}{$-847.5100$} & --  & --   \\
		                     & \multicolumn{1}{l}{$-847.5075$} & 1.81  & 4.4827   \\
		                     & \multicolumn{1}{l}{$-847.5025$} & 3.85  & 4.0901   \\
		                     & \multicolumn{1}{l}{$-847.4950$} & 5.70  & 3.8021   \\ \hline
		\multirow{4}{*}{F24} & \multicolumn{1}{l}{$-847.5100$} & --  & --   \\
		                     & \multicolumn{1}{l}{$-847.5075$} & 1.00  & 4.6668   \\
		                     & \multicolumn{1}{l}{$-847.5025$} & 3.15  & 4.1665   \\
		                     & \multicolumn{1}{l}{$-847.4950$} & 5.47  & 3.7926   \\ \hline
		\multirow{4}{*}{F32} & \multicolumn{1}{l}{$-847.5100$} & 0.88  & 6.4009   \\
		                     & \multicolumn{1}{l}{$-847.5075$} & 2.61  & 5.7128   \\
		                     & \multicolumn{1}{l}{$-847.5025$} & 4.57  & 5.4137   \\
		                     & \multicolumn{1}{l}{$-847.4950$} & 2.63  & 5.2941  \\ 
                       \hline

	\end{tabular}
	
\end{table}

\subsection{Case III: $\text{225}{\text{ }}\text{deg} $}\label{subsec4.3}
\begin{figure}[!htb]%
\centering
\includegraphics[width=0.8\textwidth]{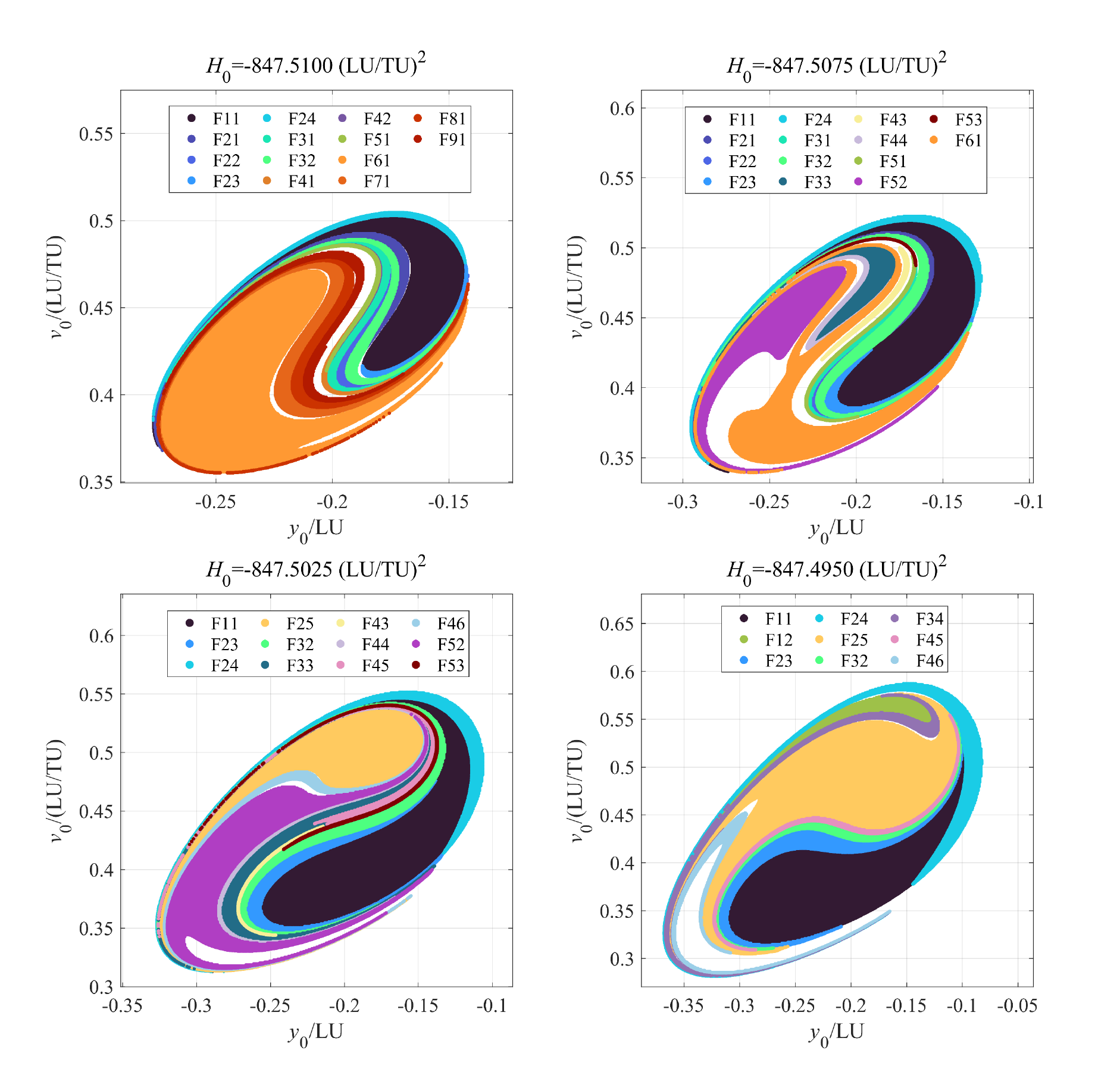}
\caption{Initial state sets of transit orbit families under four $H_0$ values.}\label{fig27}
\end{figure}

When $\theta _{{\text{S0}}}$ differs by 180 deg from ${\theta _{{\text{S0}}}} = 45{\text{ }}\deg $, the dynamical phenomenon appears to be quite similar. In Fig. \ref{fig27}, it is found that the classifications of transit orbits at ${\theta _{{\text{S0}}}} = 225{\text{ }}\deg $ are almost identical to those at ${\theta _{{\text{S0}}}} = 45{\text{ }}\deg $ under the same $H_0$. The dynamical behaviors and transfer characteristics of transit orbit families are similar at these two $\theta _{{\text{S0}}}$ values. This phenomenon also exists for any two $\theta _{{\text{S0}}}$ values that differ by 180 deg \citep{bib9}. Considering the energy conditions required for the existence of the LCSs at each $\theta _{{\text{S0}}}$ shown in Table \ref{tab2}, it confirms the aforementioned conclusions that $\theta _{{\text{S0}}}$ affects the energy conditions of emergence and disappearance of transit orbit families by affecting the configurations of the LCSs under different $H_0$. This finding expands on the results from Ren and Shan \citep{bib9}, and our future research will further investigate this phenomenon.

\section{Application of classifications on low-energy transfer}\label{sec5}
In this section, the classifications and their evolution laws presented in Section \ref{sec4} are applied to the constructions of two low-energy transfer scenarios, i.e., bi-impulsive Earth-Moon transfer (without and with LF \citep{bib10,bib47,bib48}) and cislunar escape. The construction strategies of these two scenarios are proposed based on the classifications and their evolution laws. Samples of transfer trajectories are presented and the link between classifications, evolution laws, and transfer characteristics is revealed.
\subsection{Bi-impulsive Earth-Moon transfer}\label{subsec5.1}
Bi-impulsive Earth-Moon transfer is described as a process in which the test body (spacecraft) is launched to a transfer trajectory from a circular Earth parking orbit after performing an Earth injection impulse ($\Delta {v_i}$) and inserted into a circular lunar insertion orbit through a Moon insertion impulse ($\Delta {v_f}$). To maximize energy variations, the impulses should be tangential to the orbital velocity \citep{bib49}. Therefore, the states of departure point at the Earth parking orbit and insertion point at the lunar insertion orbit should satisfy \citep{bib8,bib11}:
\begin{equation}
{\bm{\psi }_i} = \left[ {\begin{array}{*{20}{c}}
  {{{\left( {{x_i} + \mu } \right)}^2} + {y_i}^2 - {{\left( {{R_{\text{E}}} + {h_i}} \right)}^2}} \\ 
  {\left( {{x_i} + \mu } \right)\left( {{u_i} - {y_i}} \right) + {y_i}\left( {{v_i} + {x_i} + \mu } \right)} 
\end{array}} \right] = \mathbf{0} \label{eq16}
\end{equation}
\begin{equation}
{\bm{\psi }_f} = \left[ {\begin{array}{*{20}{c}}
  {{{\left( {{x_f} + \mu  - 1} \right)}^2} + {y_f}^2 - {{\left( {{R_{\text{M}}} + {h_f}} \right)}^2}} \\ 
  {\left( {{x_f} + \mu  - 1} \right)\left( {{u_f} - {y_f}} \right) + {y_f}\left( {{v_f} + {x_f} + \mu  - 1} \right)} 
\end{array}} \right] = \mathbf{0} \label{eq17}
\end{equation}
where $h_i$ denotes the altitude of the Earth parking orbit and $h_f$ denotes the altitude of the lunar insertion orbit. The subscript ‘\textit{i}’ denotes quantities associated with the departure point, while ‘\textit{f}’ denotes quantities associated with the insertion point. In this paper, $h_i$ is set to 36000 km and $h_f$ is set to 100 km. Bi-impulsive Earth-Moon transfer can be categorized into two main types, i.e., transfer without LF and transfer with LF. These two types of transfer can be constructed based on the LCSs and transit orbits with different strategies. 

In this paper, the segment of the trajectories towards the Moon (i.e., Segment I) is generated from the forward-time propagation of the initial states of transit orbits inside the LCSs, the segment of the trajectories departing from the Earth parking orbits (i.e., Segment II) is generated from the backward-time propagation. For transfer without LF, part of the transit orbits consists of Segment I (i.e., the propagation time of Segment I $T_{1}<T_\text{Passage}$), with a periapsis satisfying ${h_p} = 100{\text{ km}}$ for the lunar insertion orbit. For transfer with LF, transit orbits play the role of lunar flyby trajectories in Segment I, and the orbital propagation continues (i.e., $T_1>T_\text{Passage}$) when the transit orbits reach $U_1$ or $U_2$. Note that trajectories inserted into the surface of the Earth or Moon are excluded in the construction. Segment I and Segment II are searched from $M_3$ inside the LCSs to satisfy the constraints \eqref{eq16}-\eqref{eq17}. When the transfers are constructed, the Earth injection impulse ($\Delta {v_i}$), the Moon insertion impulse ($\Delta {v_f}$), and the total impulse ($\Delta {v}$) are calculated by:
\begin{equation}
\Delta {v_i} = \sqrt {{{\left( {{u_i} - {y_i}} \right)}^2} + {{\left( {{v_i} + {x_i} + \mu } \right)}^2}}  - \sqrt {\frac{{1 - \mu }}{{{R_{\text{E}}} + {h_i}}}}  \label{eq18}
\end{equation}
\begin{equation}
\Delta {v_f} = \sqrt {{{\left( {{u_f} - {y_f}} \right)}^2} + {{\left( {{v_f} + {x_f} + \mu  - 1} \right)}^2}}  - \sqrt {\frac{\mu }{{{R_{\text{M}}} + {h_f}}}}   \label{eq19}
\end{equation}
\begin{equation}
\Delta v = \Delta {v_i} + \Delta {v_f}  \label{eq20}
\end{equation}
\subsubsection{Bi-impulsive Earth-Moon transfer without LF}\label{subsubsec5.1.1}

\begin{figure}[!htb]%
\centering
\includegraphics[width=0.8\textwidth]{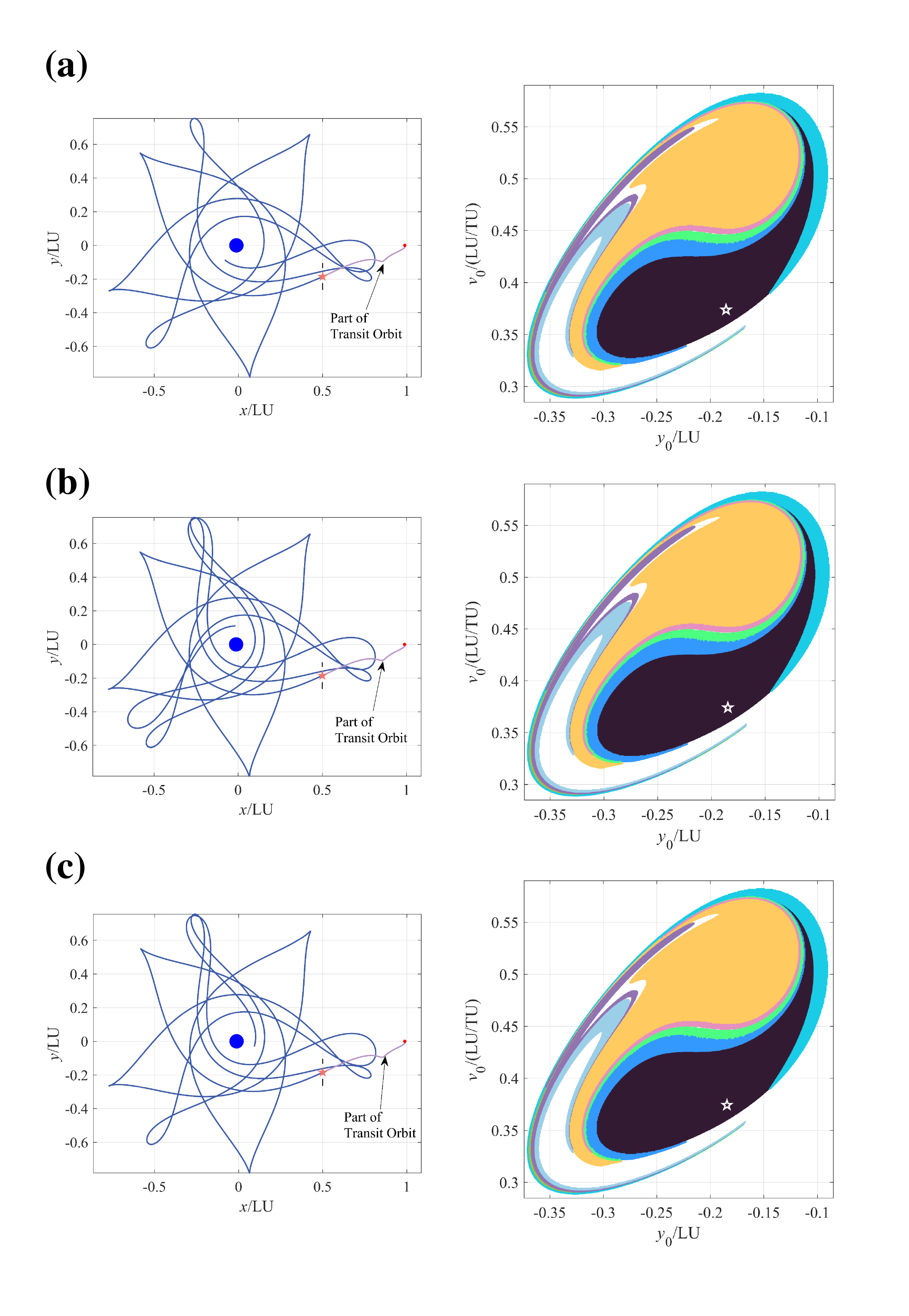}
\caption{Samples of bi-impulsive Earth-Moon transfers without LF. (a) Sample I; (b) Sample II; (c) Sample III.}\label{fig28}
\end{figure}

Based on the aforementioned discussions on the ranges of $h_p$ in Section 4, we select transit orbits with periapses satisfying ${h_p} = 100{\text{ km}}$ for the lunar insertion orbit to construct the transfers (i.e., Strategy I). For example, the transit orbits associated with family F11 under $\left( {135\text{ }\deg , - 847.4950{\text{ }}{{({\text{LU/TU}})}^2}} \right)$ (${h_p} \in \left[ { - 1.7061 \times {{10}^3},1.8823 \times {{10}^4}} \right]{\text{ }} {{\text{km}}}$) are selected. Figure \ref{fig28} presents three samples of transfers and the initial states inside the corresponding LCS. In these three samples, transit orbits insert into lunar insertion orbit at the first periapsis, validating the effectiveness of the proposed strategy.
\subsubsection{Bi-impulsive Earth-Moon transfer with LF}\label{subsubsec5.1.2}

\begin{figure}[!htb]%
\centering
\includegraphics[width=0.8\textwidth]{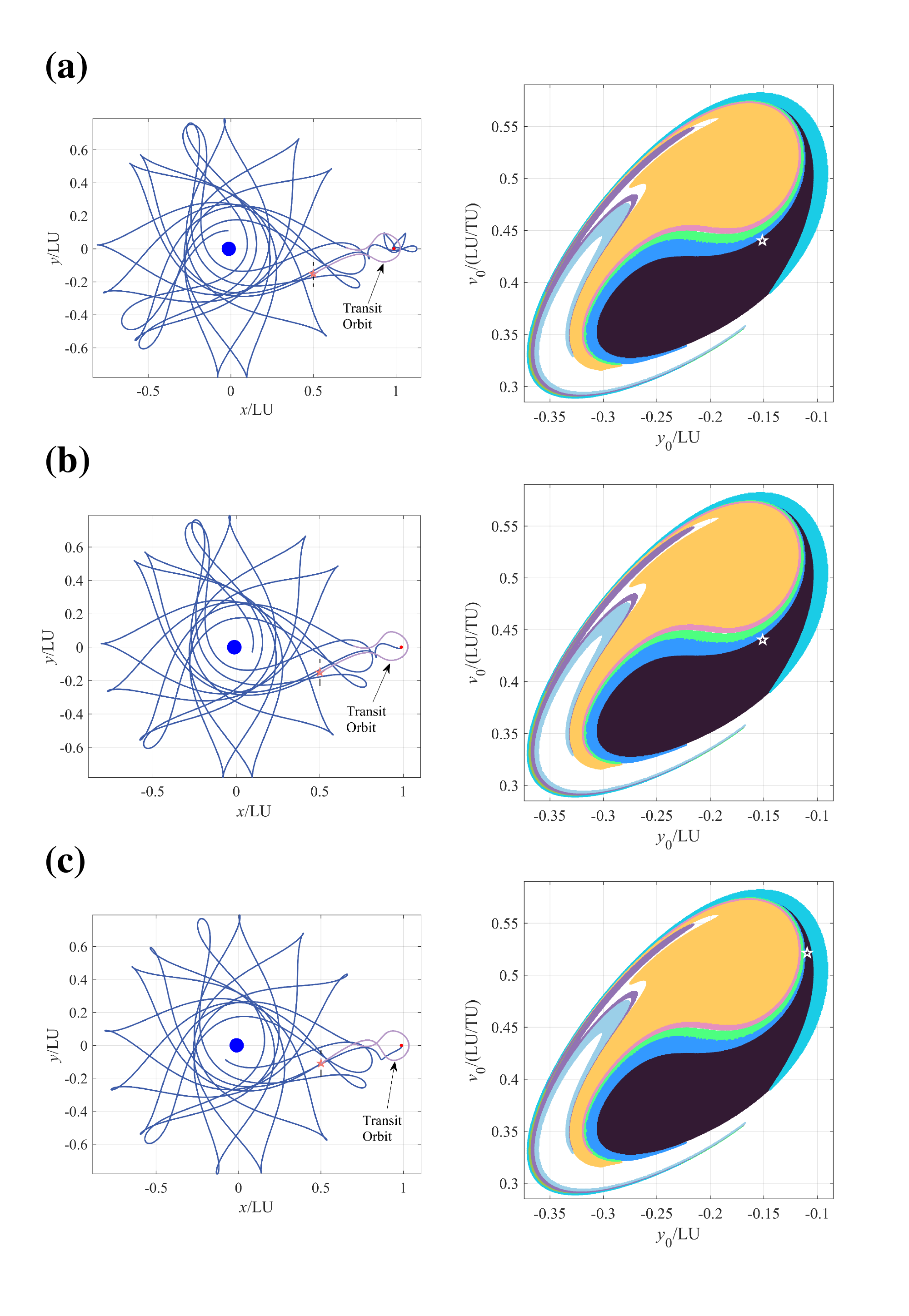}
\caption{Samples of bi-impulsive Earth-Moon transfers with LF. (a) Sample I; (b) Sample II; (c) Sample III.}\label{fig29}
\end{figure}

For transfer with LF, transit orbits consist of lunar flyby trajectories in Segment I. For this consideration, the trajectory patterns of families F11 and F12 are suitable due to their relatively short $\left\langle {{T_{{\text{Passage}}}}} \right\rangle $ (i.e., Strategy II). Transit orbits associated with family F11 under $\left( {135\text{ }\deg , - 847.4950{\text{ }}{{({\text{LU/TU}})}^2}} \right)$ are selected as an example. Figure \ref{fig29} presents three samples of transfers. Trajectories encounter the Moon twice during the transfers. In each sample, the first encounter with the Moon is an LF based on the transit orbit associated with family F11, while the second encounter is an insertion to the lunar insertion orbit.

Table \ref{tab5} presents the impulses and time of flight (TOF) of the aforementioned samples compared with the results obtained from the patched-conic method based on the patched restricted two-body problem (R2BP) (i.e., Solution I) \citep{bib50} and theoretical minimum impulse estimation (i.e., Solution II) \citep{bib51}  based on the Earth-Moon PCR3BP. It is concluded that while the TOFs of transfers with LF are longer than those without LF, the samples with LF have the advantage of requiring relatively lower impulses. Comparing these six samples with the patched-conic results, a significant reduction in impulses is achieved. Our results are closer to the theoretical minimum estimation, indicating that the construction strategies based on the classifications for low-energy transfers are effective.
\begin{table}[h]
\centering
\renewcommand{\arraystretch}{1.5}
\caption{Comparison of Transfer characteristics}\label{tab5}%
\begin{tabular}{@{}lllll@{}}
\hline
Solutions & $\Delta {v_i}/({\text{km/s}})$  & $\Delta {v_f}/({\text{km/s}})$ & $\Delta {v}/({\text{km/s}})$ & TOF/day\\
\hline
Sample I\footnotemark[1]   & 0.9937   & 0.6357  & 1.6294 & 92  \\
Sample II\footnotemark[1]    & 0.9949   & 0.6357  & 1.6306 & 109  \\
Sample III\footnotemark[1]    & 0.9944   & 0.6357  & 1.6301  & 101 \\
Sample I\footnotemark[2]   & 0.9946   & 0.6294  & 1.6240 & 218  \\
Sample II\footnotemark[2]    & 0.9943   & 0.6296  & 1.6239 & 196  \\
Sample III\footnotemark[2]    & 0.9936   & 0.6336  & 1.6272  & 185 \\
Solution I   & 1.0511   & 0.7094  & 1.7605 & 6  \\
Solution II   & 0.9548   & 0.6250  & 1.5798 & --  \\
\hline
\end{tabular}
\footnotetext[1]{Tranfers without LF.}
\footnotetext[2]{Tranfers with LF.}
\end{table}

\subsection{Cislunar escape}\label{subsec5.2}
Cislunar escape trajectories are those departing from a circular Earth parking orbit and entering the Earth-Moon exterior region. Similar to the bi-impulsive Earth-Moon transfer scenario, the trajectories departing from the Earth parking orbits (Segment II) and the trajectories escaping from cislunar space (Segment I, $T_1>T_\text{Passage}$) form cislunar escape trajectories. The test body (spacecraft) requires a tangential Earth injection impulse ($\Delta {v_i}$) to depart. The altitude of circular Earth parking orbit $h_i$ is also set to 36000 km. Therefore, the states of departure point at the Earth parking orbit should satisfy Eq. \eqref{eq16}, while $\Delta {v_i}$ can be calculated by Eq. \eqref{eq18}. The numerical criteria for escape is established such that when the states $\bm{X}$ satisfy $\sqrt {{x^2} + {y^2}}  > 10{\text{ LU}}$, the trajectories are considered as escape trajectories \citep{bib15}. 

\begin{figure}[!htb]%
\centering
\includegraphics[width=1.0\textwidth]{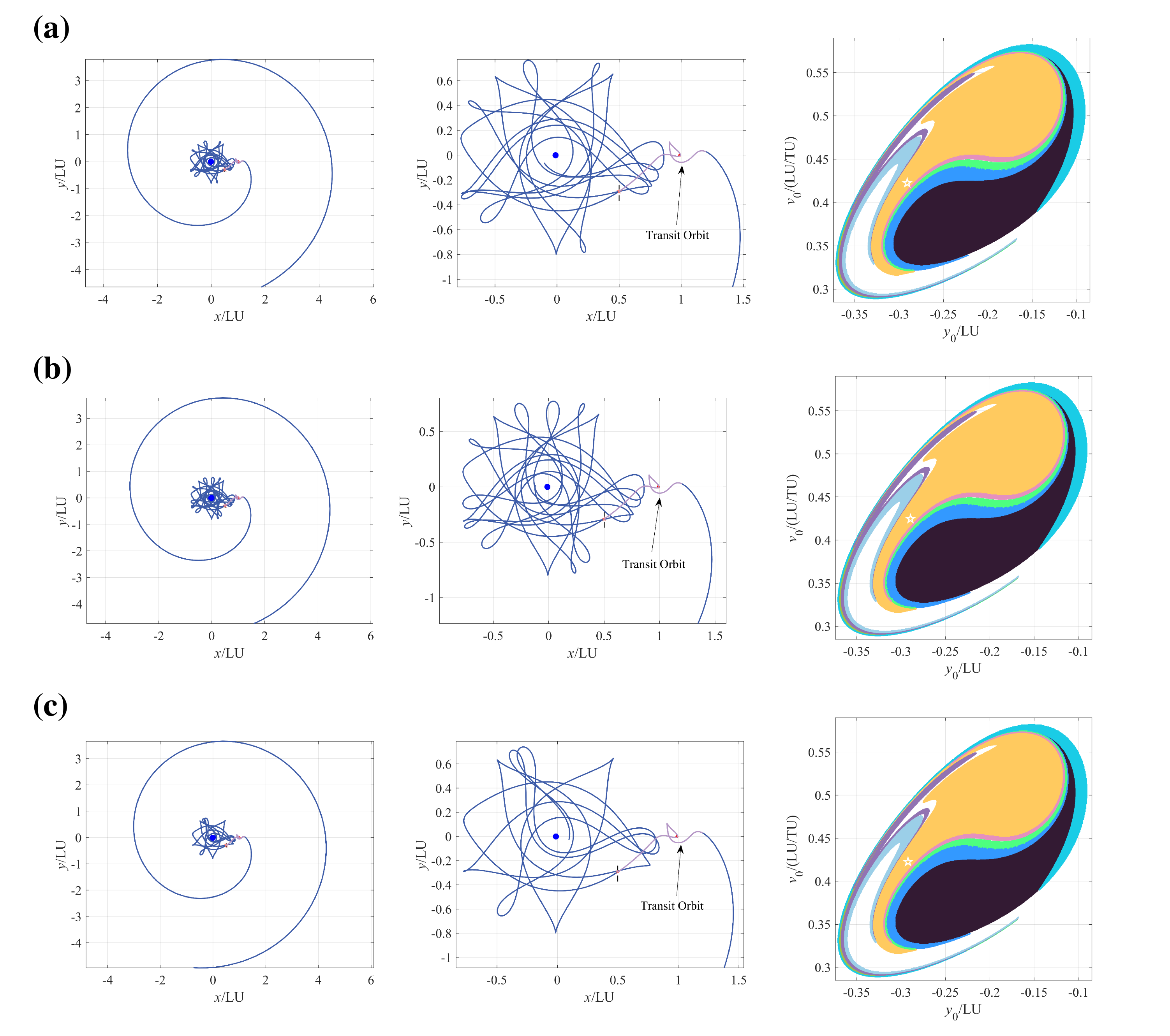}
\caption{Samples of cislunar escape trajectories. (a) Sample I; (b) Sample II; (c) Sample III.}\label{fig30}
\end{figure}

As mentioned in Sections \ref{sec3} and \ref{sec4}, L2 escape trajectories pass through the L2 region and enter the Earth-Moon exterior region. Therefore, L2 escape trajectories are applied to constructing cislunar escape trajectories (i.e., Strategy III). For example, transit orbits associated with family F25 under $\left( {135\text{ }\deg , - 847.4950{\text{ }}{{({\text{LU/TU}})}^2}} \right)$ are selected. Figure \ref{fig30} presents three samples (note that parts of escape trajectories are shown for clarity). It is found that transit orbits connect the trajectories in the Earth region (i.e., Segment II) with the trajectories in the Earth-Moon exterior region.

A comparison between the $\Delta {v_i}$ of these three samples, escape trajectories based on the R2BP (i.e., Solution I) \citep{bib52}, and theoretical minimum impulse estimation (i.e., Solution II) based on the Earth-Moon PCR3BP is shown in Table \ref{tab6}. The calculation methods of $\Delta {v_i}$ of Solution I and Solution II are detailed in Appendix B. It is found that three samples advantage in low $\Delta {v_i}$ compared with Solution I. The low-energy escape is achieved by using natural multi-body dynamics, and our proposed strategy links the two based on the classifications of transit orbits.
\begin{table}[h]
\centering
\renewcommand{\arraystretch}{1.5}
\caption{Comparison of transfer characteristics}\label{tab6}%
\begin{tabular}{@{}llllll@{}}
\hline
Solutions    & Sample I   & Sample II  & Sample III  & Solution I & Solution II  \\
\hline

$\Delta {v_i}/({\text{km/s}})$    &  0.9914  & 0.9882  & 0.9944  & 1.2687 & 0.9569  \\
\hline
\end{tabular}

\end{table}

\section{Conclusion}\label{sec6}
 
This paper focuses on the classification of interior transit orbits in the Sun-Earth/Moon planar bicircular restricted four-body problem (PBCR4BP). Lagrangian coherent structures (LCSs) are introduced to generate the transit orbits and to visualize the classification. The number of periapses about the Moon is selected as the classification parameter. Utilizing the number of periapses, a clear classification boundary has been achieved, and the classification maps have been presented. The evolution laws of the classifications with respect to the initial Hamiltonian $H_0$ and the initial solar phase angle ${\theta _{{\rm{S0}}}}$ have been discussed and analyzed. It is concluded that $H_0$ affects the transfer characteristic parameters and the emergence and disappearance of transit orbit families, while ${\theta _{{\rm{S0}}}}$ affects the energy conditions of the emergence and disappearance of the families by affecting the configurations of the LCSs under different $H_0$.  Based on the classifications and their evolution laws, three construction strategies of low-energy transfers (i.e., bi-impulsive Earth-Moon transfer and cislunar escape) are proposed. Numerical simulations of the transfer trajectories verify the effectiveness of the proposed strategies, and a direct link between classifications, evolution laws, and transfer characteristics is finally established.

\section*{Acknowledgements}

The first author acknowledges the financial support from the Outstanding Research Project of Shen Yuan Honors College, BUAA (Grant No. 230122205). The second author acknowledges the financial support from the National Natural Science Foundation of China (Grant No. 12302058). The third author acknowledges the financial support from the National Natural Science Foundation of China (Grant No. 12372044).

\begin{appendices}

\section{Details about family F50}\label{secA1}

\begin{figure}[!htb]%
\centering
\includegraphics[width=1.0\textwidth]{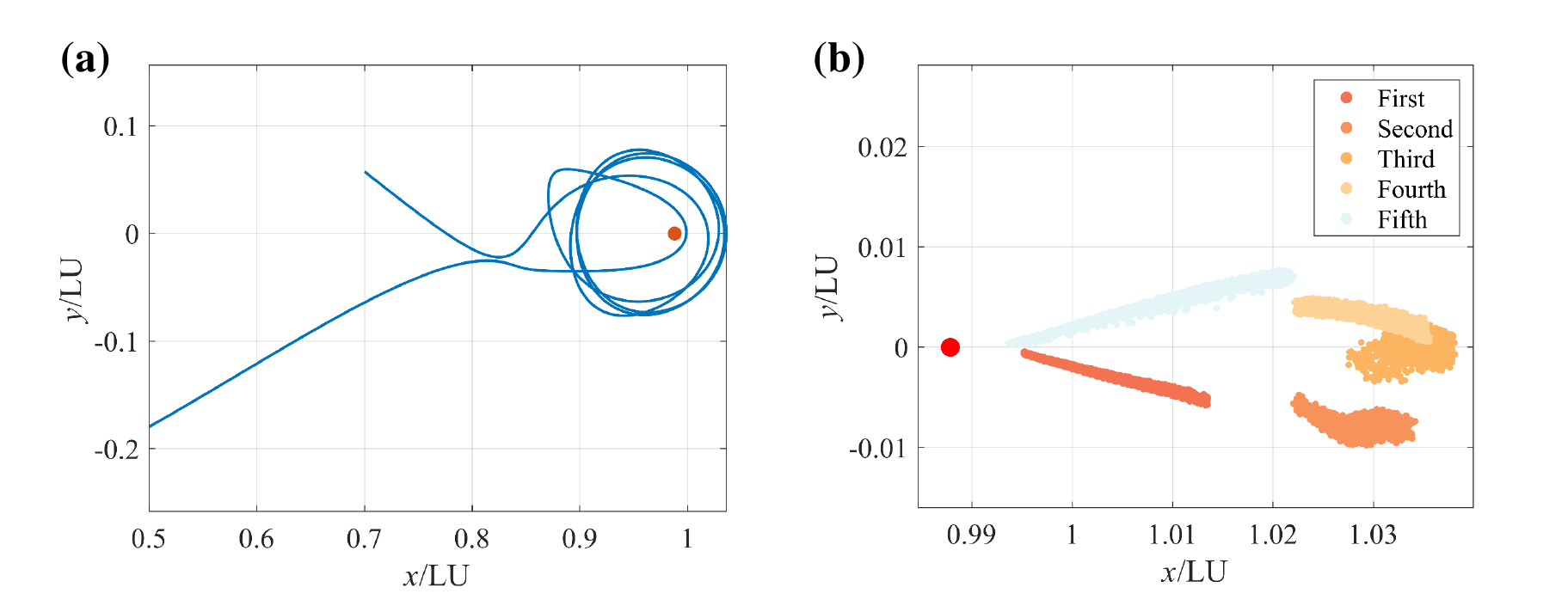}
\caption{Samples of bi-impulsive Earth-Moon transfers with LF. (a) Typical trajectory pattern; (b) Periapsis distribution.}\label{figA1}
\end{figure}

The typical pattern of transit orbits associated with family F50 and the periapsis distribution are presented in Fig. \ref{figA1}. It is found that transit orbits associated with family F50 are short-term capture trajectories.
\section{Calculation methods of $\Delta {v_i}$ of escape trajectories}\label{secB1}
For Solution I shown in Table \ref{tab6}, the parabola is the escape trajectory with the minimum energy \citep{bib52}. The velocity at the perigee (i.e., the departure point) is calculated by:
\begin{equation}
 v_1 = \sqrt {\frac{{2(1 - \mu) }}{{{R_{\text{E}}} + {h_i}}}}\label{eqB1}
\end{equation}
Then, $\Delta {v_i}$ is calculated by:
\begin{equation}
\Delta {v_i} =  v_1 - \sqrt {\frac{{1 - \mu }}{{{R_{\text{E}}} + {h_i}}}} = (\sqrt{2}-1) \sqrt {\frac{{1 - \mu }}{{{R_{\text{E}}} + {h_i}}}}\label{eqB2}
\end{equation}
For Solution II, the theoretical minimum impulse is estimated by the difference of Jacobi energy (\textit{C}) between the Earth parking orbit and the L2 point in the Earth-Moon circular restricted three-body problem. When the Jacobi energy of the trajectory is the Jacobi energy of L2 $C_\text{L2}$, the trajectory is capable of escape. The Jacobi energy is expressed as \citep{bib27,bib49}:
\begin{equation}
C = -\left( {{u^2} + {v^2}} \right) +  \left( {{x^2} + {y^2}} \right) + \frac{{2(1 - \mu) }}{{{r_1}}} + \frac{2\mu }{{{r_2}}} + \mu \left( {1 - \mu } \right) \label{eqB3}
\end{equation}
Then, the theoretical minimum impulse is estimated by:
\begin{equation}
\Delta {v_i} =  \sqrt{\delta{C}+\frac{{1 - \mu }}{{{R_{\text{E}}} + {h_i}}}} - \sqrt {\frac{{1 - \mu }}{{{R_{\text{E}}} + {h_i}}}}\label{eqB4}
\end{equation}
where $\delta{C}$ denotes the difference of \textit{C} between the Earth parking orbit and L2 ($C_\text{L2}=3.184158{\text{ }}{({\text{LU/TU}})^2}$ in the Earth-Moon CR3BP).



\end{appendices}







\end{document}